\documentclass[aps,pra,reprint,showpacs,floatfix]{revtex4-1}

\usepackage{amsmath}
\usepackage{slashed}
\usepackage{txfonts}
\usepackage{microtype}
\usepackage{multirow}
\usepackage{graphicx}
\usepackage[
breaklinks=true
]{hyperref}
\usepackage{color}
\usepackage{isomath}
\def\sout{\bgroup\markoverwith
{\textcolor{red}{\rule[0.5ex]{2pt}{0.5pt}}}\ULon}
\def\be{\begin{equation}}
\def\ee{\end{equation}}
\def\bes{\begin{equation*}}
\def\ees{\end{equation*}}
\def\bea{\begin{eqnarray}}
\def\eea{\end{eqnarray}}
\def\beas{\begin{eqnarray*}}
\def\eeas{\end{eqnarray*}}
\def\bal#1\eal{\begin{align}#1\end{align}}
\def\bals#1\eals{\begin{align*}#1\end{align*}}
\newcommand{\bra}[1]{\langle #1|}
\newcommand{\ket}[1]{|#1\rangle}

\renewcommand{\vec}{\vectorsym}
\newcommand{\del}{\partial}

\graphicspath{{figures/}}

\renewcommand*{\vec}[1]{\boldsymbol{#1}}

\begin{document}
\title{Above-threshold ionization with highly charged ions in super-strong laser 
fields:\\ III. 
Spin effects and its dependence on laser polarization}

\author{Enderalp \surname{ Yakaboylu}}\thanks{\mbox{enderalp.yakaboylu@mpi-hd.mpg.de }}
\affiliation{Max-Planck-Institut f\"ur Kernphysik, Saupfercheckweg 1, D-69117 
Heidelberg, Germany}
\author{Michael \surname{Klaiber}}\thanks{\mbox{michael.klaiber@mpi-hd.mpg.de }}
\affiliation{Max-Planck-Institut f\"ur Kernphysik, Saupfercheckweg 1, D-69117 
Heidelberg, Germany}
\author{Karen Z. \surname{Hatsagortsyan}}\thanks{\mbox{k.hatsagortsyan@mpi-hd.mpg.de }}
\affiliation{Max-Planck-Institut f\"ur Kernphysik, Saupfercheckweg 1, D-69117 
Heidelberg, Germany}

\date{\today}

\begin{abstract}

Spin effects in the tunneling regime of strong field ionization of hydrogenlike highly charged ions in linearly as well as circularly polarized laser fields are investigated. The impact of the polarization of a laser field on the spin effects are analyzed. Spin-resolved differential ionization rates are calculated employing the relativistic Coulomb-corrected strong-field approximation (SFA) developed in the previous paper of the series. Analytical expressions for spin asymmetries and spin flip probability,  depending on the laser's polarization, are obtained for the photoelectron momentum corresponding to the maximum of tunneling probability. A simpleman model is developed for the description of spin dynamics in tunnel-ionization, which provides an intuitive explanation for the spin effects. The  spin flip is shown to be experimentally observable by using  moderate highly charged ions with a charge of the order of 20 and a laser field with an intensity of $I\sim 10^{22}$ W/cm$^2$.

\end{abstract}

\maketitle

\section{Introduction} \label{sec:introduction}

Due to advances of the laser technology, the relativistic regime of laser-atom interactions is now within the experimental reach 
\cite{Moore_1999,Chowdhury_2001,Dammasch_2001,Yamakawa_2003,Gubbini_2005,DiChiara_2008,Palaniyappan_2008,DiChiara_2010,Ekanayake_2013,RMP_2012}.
Recently different tools for advanced studies of particles in extreme laser fields have been also developed, which make it possible to select and prepare well-defined ion ensembles and to optimize the laser-particle interaction \cite{Vogel_2012}. This will allow the investigation of the relativistic strong field ionization dynamics, which is accessible only in a combination of strong lasers and highly charged ions.

The strong field approximation (SFA) \cite{Reiss_1990,Reiss_1990b} and the imaginary-time method (ITM) \cite{Popov_1997,Mur_1998,Popov_2004,Milosevic_2002r1,Milosevic_2002r2}
are well developed theoretical tools for the analytical investigation of strong field ionization in the relativistic regime. In particular, the differential and total ionization rates were calculated using these methods. The photoelectron momentum distribution in the relativistic regime of above-threshold ionization was well explained by those calculations \cite{Krainov_1992,Krainov_1995,Krainov_1999,Ortner_2000a,Ortner_2000b,Krainov_2008,Becker_2002}. However, the details of the electron spin dynamics in the relativistic tunnel-ionization regime still needs further elucidation.

Spin effects in different processes in laser fields have been investigated since the invention of the laser \cite{Bunkin_1972}. In particular, it was shown that the relativistic dynamics of free electrons in a strong laser field is disturbed by spin induced forces \cite{Walser_1999,Wen_2014}, and that the electron radiation can be modified due to the spin induced dynamics \cite{Walser_2000b}. Spin effects in laser assisted Mott scattering and laser assisted M\"oller scattering were investigated in 
\cite{Szymanowski_1997,Panek_2002} and  \cite{Panek_2004}, respectively.  Polarization effects in the multiphoton Compton scattering were investigated 
in \cite{Ternov_1968,Goldman_1969,Bagrov_1989,Bolshedvorsky_2000,Kotkin_2003,Kotkin_2005,Karlovets_2011,Boca_2012,Krajewska_2013}. Transfer of polarization from the laser beam to positrons via Compton scattering and pair production is shown in \cite{Omori_2006}. Recently, a spin flip effect was shown in the Kapitza-Dirac effect \cite{Ahrens_2012,Ahrens_2013,Sven2013spin}. Relativistic spin operators in various electromagnetic environments have been discussed in \cite{Bauke_2014,Bauke_2014b}. 
Collapse and revival of the spin precession in a laser field has been revealed in \cite{Skoromnik_2013}. Furthermore, interesting polarization effects were explored in electron-positron pair production processes in ultra-strong laser fields, in particular, during electron and positron pair creation in combined Coulomb and strong laser fields  
\cite{DiPiazza_2010}, in the multiphoton Bethe-Heitler process \cite{Muller_2012} and in two counterpropagating laser pulses \cite{Woellert_2015}. Spin correlations in electron-positron pair creation by a laser pulse and a proton beam was examined in \cite{Muller_2011}.

During the relativistic laser-atom interaction spin effects were shown to appear in the laser-driven bound electron dynamics \cite{Walser_2002} and, in particular, in the radiation of high-order harmonics \cite{Hu_1999,Hu_2001,Walser_2001,Keitel2001relativistic}. The spin asymmetry in the strong-field-ionization process of an atom with a circularly polarized laser field was discussed in \cite{Faisal_2004}, neglecting the spin dynamics in the bound state. The latter, however, can have a significant impact on spin effect as it was shown in \cite{Klaiber2014spin}. The spin dynamics in nonsequential double ionization of helium was considered in \cite{Bhattacharyya_2007,Bhattacharyya_2011}. Furthermore, the photoelectron spin polarization can also arise because of the electron-ion entanglement \cite{Barth_2013,Barth_2014}. The spin-asymmetry in the relativistic regime of tunnel-ionization from $p$-states was explored in \cite{Klaiber_2014p}.

In this paper we investigate the dependence of the spin effects during tunnel-ionization on the polarization of the laser field. The spin-resolved differential ionization rates of a highly charged hydrogenlike ion from the ground  state in a strong laser field of linear and circular polarizations are calculated using the relativistic Coulomb-corrected SFA (cc-SFA) developed in paper II (the second paper of this series \cite{Klaiber_2013a,Klaiber_2013b}). Spin asymmetries and spin-flip effect during direct ionization of an hydrogenlike system are investigated for the peak of the final momentum distribution. Similarities and differences of the spin asymmetries and spin-flip effect in the cases of linear and circular polarizations of a laser field are analyzed. In addition to the standard relativistic Coulomb corrected SFA (s-cc-SFA), we also apply 
a dressed cc-SFA (d-cc-SFA), which is based on the use of a non-standard partition of the Hamiltonian within the SFA formalism \cite{Faisal_2007a,Faisal_2007b}. The physical relevance of the different versions of the SFA formalism is discussed. While in s-cc-SFA the influence of the laser field on the electron spin evolution in the bound state is not taken into account, it is fully accounted for in d-cc-SFA, which is shown to have a decisive impact on the spin effects. Finally, we provide a simpleman model for intuitive understanding of the spin effects. It  incorporates the propagators of the spin states for the bound and continuum motion, and quasiclassical description of the tunneling process.

Spin effects in the tunneling regime of ionization emerge through three steps \cite{Klaiber2014spin}; spin precession in the bound state, spin rotation during tunneling, and spin precession during the electron motion in the continuum. In s-cc-SFA the spin dynamics in the bound state is completely neglected. Therefore, in this case the spin effects are determined by the electron dynamics during tunneling and the motion in the continuum. Because of the evident asymmetry in the spin evolution in this picture, relatively large spin effects arise. However, it is known that the laser field can induce a large spin precession in the bound state \cite{Hu_1999,Hu_2001}, which is accounted for in d-cc-SFA \cite{Klaiber_2013b,Klaiber2014spin}. It reduces the asymmetry in the spin dynamics during ionization and, consequently, lead to a reduction of spin effects. Analysis based on our simpleman model shows that the spin asymmetries are a consequence of the tunneling step.

The plan of the paper is the following. The spin-resolved differential ionization rates within the two versions of cc-SFA (standard and dressed) are calculated in Sec.~\ref{sec:cc_SFA}. The bound state spin dynamics, which is essential for d-cc-SFA is investigated in Sec.~\ref{sec:bound_state}. The final momentum distribution of the tunnel-ionized electron is presented in Sec.~\ref{mom_at_saddle}, which later is used in the calculation of the spin effects for the maximal tunneling probability. Analytical formulas for the spin asymmetries and the spin flip in linearly and circularly polarized laser fields are calculated in Sec.~\ref{sec:spin_asymmetries}. A simpleman model for the spin dynamics is presented in Sec.\ref{sec:intuitive}. The possibilities for an experimental observation of spin effects are discussed in Sec.~\ref{sec:experiment}. Our conclusion is given in Sec.~\ref{sec:conclusion}. Atomic units (a.u.) and the metric convention $g = (+,-,-,-)$ are used throughout the paper.

\section{The Coulomb corrected strong field approximation}
\label{sec:cc_SFA}

The Hamiltonian which governs the dynamics of the laser-induced tunnel-ionization from the ground state of a hydrogenlike ion, is given by
\be
H = c \vec{\alpha} \cdot (\vec{p} + \vec{A}_L) - A^0_L + \beta c^2  + V(r) \, ,
\ee
where the laser field is described by gauge potentials given in G\"{o}ppert-Mayer gauge as $A_L^\mu  = (A^0_L,c\vec{A}_L) = - \vec{x} \cdot \vec{E}(\eta) (1, \hat{\vec{k}})$, with the wave vector $k^\mu = \omega /c (1, \hat{\vec{k}})$, the laser frequency $\omega$ and the phase $\eta = k x /\omega $, and the Coulomb potential $V(r) = - \kappa / r$, $\kappa$ being the charge of the hydrogenlike ion, and $\beta,\,\vec{\alpha}$ are the Dirac matrices.

The transition amplitude between the initial state $\ket{\Psi_i^s}$ with the magnetic spin quantum number $s$ and the final state $\ket{\Psi_f^{s'}}$ with the number $s'$ can be written as 
\be
\label{S-1}
M_{s \rightarrow s'} = (S-1)_{s \rightarrow s'} = - i \int_{-\infty}^\infty dt \, \bra{\Psi_f^{s'}} H_{int}
\ket{\Psi_i^s} \, ,
\ee
with the interaction Hamiltonian~$H_{int}$. The S-matrix treatment is exact as far as it incorporates the exact final state~$\ket{\Psi_f^{s'}}$, which is the exact solution of the Schr\"odinger equation 
\be
\label{psi_1}
i \partial_t \ket{\Psi_f^{s'}} = H  \ket{\Psi_f^{s'}} \, .
\ee
The initial state in its turn fulfills the following equation
\be
\label{psi_0}
i \partial_t \ket{\Psi_i^s} = (H- H_{int}) \ket{\Psi_i^s} \, .
\ee
In the SFA, the exact final state~$\ket{\Psi_f^{s'}}$ in the
transition amplitude~(\ref{S-1}) is approximated by the Volkov state \cite{Volkov_1935}, whose wave function in the G\"{o}ppert-Mayer gauge reads
\be
\label{Volkov_gmg}
\Psi_V^{s'} = \sqrt{\frac{c^2}{\varepsilon}} \exp\left( i S + i \vec{A} \cdot \vec{x} \right)
\left[1+\frac{1}{2c\lambda} 
(1+\hat{\vec{k}}\cdot \vec{\alpha}) \vec{A} \cdot \vec{\alpha} \right] v_{s'} \, ,
\ee
with  the Volkov action
\be
S(\eta) = - px - \frac{1}{\lambda} \int_{-\infty}^\eta d \eta' \, \left[\vec{A}(\eta') \cdot \vec{p} + \frac{\vec{A}(\eta')^2}{2}\right] \,
,
\ee
and the free particle spinor
\be
\label{volkov_spinor}
v_{s'} = \sqrt{\frac{\varepsilon + c^2}{2c^2}} \begin{pmatrix}
            \chi_{s'} \\
	    \\
            \displaystyle \frac{c}{\varepsilon + c^2} \vec{p} \cdot \vec{\sigma} \chi_{s'} \\
          \end{pmatrix} \, .
\ee 
Here $\varepsilon = \sqrt{c^4 + c^2 \vec{p}^2}$ is the electron energy, $\lambda = \varepsilon/c^2 - \vec{p} \cdot \hat{\vec{k}}/c $ is the integral of motion for the electron in a plane wave field, $\vec{A} \equiv -\int^\eta_{-\infty} \vec{E}(\eta') d\eta'$, and $\chi_{+} = (1 \quad 0)^T$ and $\chi_{-} = (0 \quad 1)^T$ are the two components spinors. Note that the Volkov wave function in the G\"{o}ppert-Mayer gauge is obtained by first solving the Dirac equation in the velocity gauge $A^\mu(\eta) = (0, c \vec{A}(\eta))$ and then applying a gauge transformation with the gauge function $ \vec{A}(\eta) \cdot \vec{x}$. 

In this approximation, the transition amplitude neglects the effect of the Coulomb potential on the electron dynamics in the continuum as well as the influence of the laser field on the bound state dynamics. To account for the Coulomb potential on the electron dynamics in the continuum, the relativistic cc-SFA was developed in paper II
\cite{Klaiber_2013b}. Rather than the Volkov solution for the continuum electron $\Psi_V^{s'}$, cc-SFA employs the wave function of the electron in the laser and Coulomb fields in the eikonal approximation \cite{Gersten_1975,Avetissian_1997,Krainov_1997,Avetissian_1999,Smirnova_2008}, which is given by [see Eq. (II.29)]
\be
 \Psi_C^{s'}=\Psi_V^{s'}\exp\left[ iS_c(\vec{x},\eta)\right],
 \label{Eikonal}
\ee
where
\be
 S_c(\vec{x},\eta)= \int^{\infty}_{\eta} d\eta'\frac{\varepsilon(\eta')}{c^2\lambda}V\left(\vec{x}(\eta')\right),
 \label{Eikonal_S}
\ee
with the relativistic trajectory of the electron in the laser field
$\vec{x}(\eta')=\vec{x}+\int^{\eta'}_{\eta}d\eta''\vec{p}(\eta'')/\lambda$, and the energy-momentum of the electron in the laser field
\bal
\vec{p}(\eta) &=\vec{p}+\vec{A}(\eta)+\hat{\vec{k}}\frac{\left[\vec{p}+\vec{A}(\eta)/2\right]\cdot\vec{A}(\eta)}{c\lambda},\\
 \varepsilon(\eta)&=\varepsilon+\frac{\left[\vec{p}+\vec{A}(\eta)/2\right]\cdot\vec{A}(\eta)}{\lambda} \, ,
\eal
where we have used Eqs.~(\ref{eom}) and  (\ref{eom_en}) of Appendix~\ref{ceom} and define the final values of the physical variables as $\vec{p} \equiv \vec{p}(\eta_f)$, $\varepsilon \equiv \varepsilon(\eta_f)$, and $\vec{A}(\eta_f) = 0$.

Thus, the  transition amplitude for the strong field ionization in cc-SFA reads
\be
M_{s \rightarrow s'} = - i \int_{-\infty}^\infty d t \, \bra{\Psi_C^{s'}} H_{int}
\ket{\Psi_i^s} \, .
\label{CC_S-1}
\ee
Accordingly, the spin resolved differential ionization rate for a certain spin transition can be defined as 
\be
\frac{d W_{s \rightarrow s'}}{d^3 \vec{p}} = \frac{\omega}{\pi} | M_{s \rightarrow 
s'} (\vec{p})|^2 \, ,
\ee
where the rate is averaged over a laser half-cycle.

In contrast to the exact S-matrix treatment, the results of the SFA calculation depend on the partition of the full Hamiltonian \cite{Faisal_2007a}, i.e., on the identification of the interaction Hamiltonian $H_{int}$. In the next sections we specify two different choices of interaction Hamiltonians, which yield different SFA approaches. We will calculate the spin resolved differential ionization rates in these two approaches and will later discuss their physical relevance.

We calculate the ionization rates in linearly as well as circularly polarized laser fields. In the velocity gauge, the vector potential of the laser field can be written as
\be
\label{a_field}
\vec{A}(\eta) = \frac{E_0}{\omega} \left[ \sin(\omega \eta) \hat{\vec{x}} - \zeta
\cos(\omega \eta) \hat{\vec{y}} \right] \, ,
\ee
which yields the corresponding electric and magnetic fields
\bal
\label{e_field}
\vec{E}(\eta) &= - E_0 \left[\cos(\omega \eta) \hat{\vec{x}} + \zeta\sin(\omega
\eta)
\hat{\vec{y}} \right] \, , \\
\label{b_field}
\vec{B}(\eta) &=  E_0 \left[ \zeta\sin(\omega \eta) \hat{\vec{x}} - \cos(\omega
\eta) \hat{\vec{y}} \right] \, ,
\eal
respectively, with the laser field amplitude $E_0$ and the polarization parameter $\zeta$, such that $\zeta =0$ corresponds to linear and $\zeta = 1$ is for circular polarization of the laser field. Further, we specify the propagation direction as $\hat{\vec{k}}=\hat{\vec{z}}$, which implies $\eta = t - z/c$.

\subsection{Standard cc-SFA}

In s-cc-SFA, the total Hamiltonian is partitioned as follows
\bal
H & = H_0^s + H_{int}^s \, , \\
H_0^s & = c \vec{\alpha} \cdot \vec{p} + \beta c^2  + V(r) \, , \\
H_{int}^s & = \vec{x} \cdot \vec{E} \left( 1-  \vec{\alpha} \cdot \hat{\vec{k}} \right) \, .
\eal
In this partition, the initial state, fulfilling  Eq. (\ref{psi_0}), is the ground state of a hydrogenlike ion $\ket{\psi_0^s}$, whose position representation is  
\be
\label{ground_state}
\psi_0^s (\vec{x})= \frac{\kappa^{3/2}}{\sqrt{\pi}} 
\sqrt{\frac{2-I_p/c^2}{\Gamma(3-2I_p/c^2)}} (2 \kappa
r)^{-I_p/c^2} \exp\left( -\kappa r - i \varepsilon_0 t\right) u_s \, ,
\ee
with the ground state spinor
\be
\label{ground_state_spinor}
u_s = \begin{pmatrix}
            \chi_s \\
	    \\
            \displaystyle i \frac{I_p}{c \kappa} \hat{\vec{x}} \cdot  
\vec{\sigma} \chi_s \\
          \end{pmatrix} \, ,
\ee
the ground state energy $\varepsilon_0 = c^2 - I_p $, and the ionization energy $I_p = c^2 - \sqrt{c^4 - c^2 \kappa^2}$ \cite{Bethe_1957}. Then, after plugging the eikonal-Volkov (\ref{Eikonal}) as well as the ground state ~(\ref{ground_state}) wave functions into the transition amplitude, and changing the variables from ($t,\vec{x}$) to ($\eta,\vec{x}$), we obtain 
\bal
\label{s-cc-SFA_matrix_1}
& M_{s \rightarrow s'} = N \int_{-\infty}^\infty d \eta \, e^{i \tilde{S}(\eta)} \\
\nonumber & \times \int d^3 x \,  
e^{-i
\vec{q}(\eta)\cdot \vec{x} - \kappa r+i S_c(\vec{x},\eta)} r^{-I_p/c^2}\vec{x} \cdot \vec{E}(\eta) \, v_{s'}^\dagger
(1-\vec{\alpha}\cdot \hat{\vec{k}}) u_s \, ,
\eal
with the following contracted action $\tilde{S}(\eta)$ and the relativistic kinetic momentum $\vec{q}(\eta)$:
\bal
\tilde{S}(\eta)  &= \frac{1}{\lambda} \int_{-\infty}^\eta d \eta' \left(\vec{A}(\eta') \cdot \vec{p} + \frac{\vec{A}(\eta')^2}{2} +\lambda (\varepsilon-\varepsilon_0)\right) \, ,\\
\label{mom_q}
\vec{q}(\eta) & = \vec{p} + \vec{A}(\eta) - \frac{\varepsilon-\varepsilon_0}{c} \hat{\vec{k}} \, .
\eal
Here, the prefactor is
\bal
N & = - i \sqrt{\frac{c^2}{\varepsilon}} \frac{\kappa^{3/2}}{\sqrt{\pi}}
\sqrt{\frac{2- I_p/c^2}{\Gamma(3-2I_p/c^2)}} (2 \kappa
)^{-I_p/c^2} \, . 
\eal
For the later convenience, we can also write down the contracted action in the form of
\be
\label{contracted_action}
\tilde{S}(\eta) = \frac{1}{2 \lambda} \int_{-\infty}^\eta d\eta' \left[ \vec{q}^2(\eta') 
+ \kappa^2 \right] \, .
\ee 
The $\eta$ integral in Eq.~(\ref{s-cc-SFA_matrix_1}) is performed via the saddle-point approximation (SPA). The saddle point equation, $\dot{\tilde{S}}(\eta_s)=0$, yields 
\be
\label{saddle_point_equation}
\vec{q}^2(\eta_s) + \kappa^2 = 0 \, .
\ee
As in the SPA-contour only the integration region near the saddle point makes the main contribution to the integral, the contracted action can be expanded around the saddle point $\eta_s$ as
\be
\tilde{S}(\eta) = \tilde{S}(\eta_s) + \ddot{\tilde{S}}(\eta_s) (\eta- \eta_s)^2 /2 \, .
\ee
Then, we replace $\eta$ with $\eta_s$ in the rest of the phase dependent functions, except in the term $\vec{q}(\eta)$, which after the coordinate integration yields to a singular function at the saddle point, as we will see below. Afterwards, the transition amplitude becomes
\bal
\label{s-cc-SFA_matrix_2}
& M_{s \rightarrow s'}  = N \int_{-\infty}^\infty d \eta \, \exp\left[ i \tilde{S}(\eta_s) + i \ddot{\tilde{S}}(\eta_s) (\eta- \eta_s)^2 /2 \right] \\
\nonumber & \times \int d^3 x \,  
e^{-i
\vec{q}(\eta) \cdot \vec{x} - \kappa r+i S_c(\vec{x},\eta_s)} r^{- I_p/c^2} \vec{x} \cdot \vec{E}(\eta_s) \, v_{s'}^\dagger
(1-\vec{\alpha}\cdot \hat{\vec{k}}) u_s \, ,
\eal
where the  Coulomb correction factor arises [see Eq.~(II.57)]
\bal
\label{Coulomb_corr}
\nonumber Q_r &\equiv \exp\left[i S_c(\vec{x},\eta_s)\right] \\
 & = \exp\left(\frac{2 I_p}{c^2}\right)\left(1 - \frac{I_p}{6 c^2} \right)\left(-\frac{\vec{x}\cdot \vec{E}(\eta_s)}{4 I_p} \right)^{I_p/c^2 -1} \, .
\eal
Furthermore, using the fact that $\vec{x} \cdot \hat{\vec{E}}/r \sim 1$, see \cite{Klaiber_2013b}, the transition amplitude reads,
\bal
& M_{s \rightarrow s'} = N e^{2 I_p/c^2}\left(1-\frac{I_p}{6c^2}\right)(-4 I_p)^{1- I_p/c^2} \\
\nonumber & \times \int_{-\infty}^\infty d \eta \, \frac{e^{i \tilde{S}(\eta_s) + i \ddot{\tilde{S}}(\eta_s) (\eta- \eta_s)^2 /2 }}{|E(\eta_s)|^{- I_p/c^2}}
J_0(\eta)\, v_{s'}^\dagger \left(1-\vec{\alpha}\cdot \hat{\vec{k}}\right) \tilde{u}_s (\eta)\, ,
\eal
where
\be
\tilde{u}_s (\eta) = \begin{pmatrix}
            \chi_s \\
	    \\
            \displaystyle i \frac{I_p}{c\kappa} \frac{\vec{J}_1 (\eta) \cdot 
\vec{\sigma}}{J_0 (\eta)} \chi_s \\
          \end{pmatrix} \, \, ,
\ee
with
\bal
J_0(\eta) & = \int d^3 x \, \exp\left[ - i \vec{q}(\eta) \cdot \vec{x}  -\kappa r \right] \, 
, \\
\vec{J}_1(\eta) \cdot \vec{\sigma} & = \int d^3 x \, \exp\left[ - i \vec{q}(\eta) \cdot 
\vec{x}  -\kappa r
\right] \hat{\vec{x}} \cdot  \vec{\sigma} \, .
\eal
With the help of the plane wave expansion, these space integrals can be calculated exactly as
\bal
J_0(\eta) & = \frac{8 \pi \kappa}{\left[\vec{q}(\eta)^2 + \kappa^2\right]^2} \, ,\\
\vec{J}_1(\eta) \cdot \vec{\sigma} & = - \frac{i 8 \pi}{\left[\vec{q}(\eta)^2 + \kappa^2\right]^2} 
\vec{q}(\eta) \cdot
\vec{\sigma} \, .
\eal
Then, the transition amplitude reads
\bal
\nonumber & M_{s \rightarrow s'}= N e^{2I_p/c^2}\left(1-\frac{I_p}{6c^2}\right)(-4
I_p)^{1- I_p/c^2} 8 \pi \kappa   \\ 
 & \times \int_{-\infty}^\infty d \eta \, \frac{e^{i \tilde{S}(\eta_s) + i \ddot{\tilde{S}}(\eta_s) (\eta- \eta_s)^2 /2
}|E(\eta_s)|^{I_p/c^2}}{\left[\vec{q}(\eta)^2 + \kappa^2\right]^2} v_{s'}^\dagger \left(1-\vec{\alpha}\cdot \hat{\vec{k}}\right) \tilde{u}_s (\eta) \, .
\eal
As a final step, we can evaluate the $\eta$-integral. The pre-exponential integrand has a singularity at the saddle point because of the saddle point condition Eq.~(\ref{saddle_point_equation}). Therefore, we first expand the singular factor around the saddle point as
\be
\frac{1}{\left[\vec{q}(\eta_s)^2 + \kappa^2\right]^2}  = \frac{1}{4 \left[\vec{q}(\eta_s) \cdot 
\dot{\vec{q}}(\eta_s)\right]^2 (\eta-\eta_s)^2} \, ,
\ee
and include it in the integration following the modified SPA \cite{Gribakin_1997}. The bispinor $\tilde{u}_s(\eta)$ has no singularity at the saddle point and reads
\be
\tilde{u}_s (\eta_s) = \begin{pmatrix}
            \chi_s \\
	    \\
            \displaystyle i \frac{I_p}{c\kappa} \hat{\vec{q}}(\eta_s) \cdot 
\vec{\sigma} \chi_s \\
          \end{pmatrix} \, ,
\ee
where we have used $q(\eta_s) = i \kappa$. Finally, the transition amplitude can be written as
\bal
\label{m_matrix_cc_s_SFA}
M_{s \rightarrow s'} = \tilde{N} \frac{\exp\left(i
\tilde{S}(\eta_s)\right)}{\left[\vec{q}(\eta_s) \cdot \vec{E}(\eta_s)\right]^{3/2}} 
\frac{|E(\eta_s)|^{I_p/c^2}}{\sqrt{\lambda}} 
 v_{s'}^\dagger \left(1-\vec{\alpha}\cdot \hat{\vec{k}}\right) \tilde{u}_s (\eta_s) \, ,
\eal
with $ \tilde{N} \equiv i N  (2 \pi i)^{3/2} \kappa e^{2I_p/c^2}\left(1-\frac{I_p}{6c^2}\right)(-4
I_p)^{1- I_p/c^2}$.

\begin{figure}
  \centering
  \includegraphics[width=0.8\linewidth]{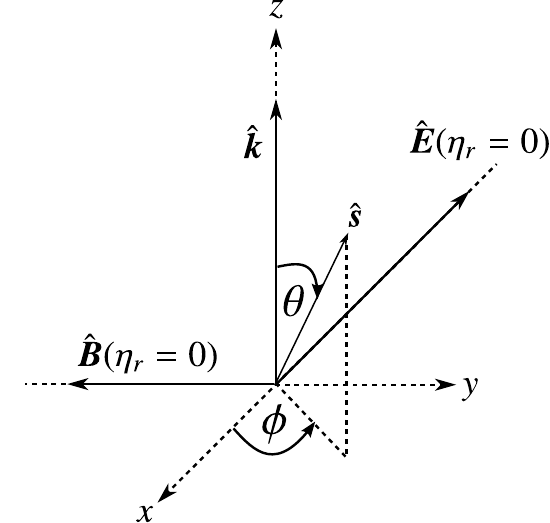}
  \caption{Spin states along an arbitrary quantization axis can be obtained by acting the rotation operator on the initial spin states whose quantization axis is along $z$. The rotation is defined by angles $(\theta, \phi)$. 
  The configuration of the laser fields at the instant of ionization is specified.}
  \label{wig_rot}
\end{figure}

In order to evaluate the transition amplitude for any spin quantization axis, we can use the rotation operator $\mathcal{D}$ after fixing the representation. Namely, we can choose the representation of the gamma matrices in the $z$-basis and specify the two-component spinor $\chi_s$ in Eq.(\ref{volkov_spinor}) as well as in Eq.(\ref{ground_state_spinor}) along the $z$-direction. Then, the spin states along an arbitrary quantization axis can be written as
\be
\ket{\Psi^s} = \sum_{s'} \mathcal{D}_{s' s} (\theta, \phi) \ket{\Psi_z^{s'}} \, ,
\ee
where $\ket{\Psi_z^s}$ is the state whose spin quantization direction is the $z$-axis and $\mathcal{D}_{s s'}(\theta,\phi)$ is the Wigner D-matrix, which can be defined as
\be
\mathcal{D}_{s' s} (\theta, \phi) = \begin{pmatrix}
            \cos(\frac{\theta}{2}) & e^{-i \phi} \sin(\frac{\theta}{2}) \\
	    & \\
            e^{i \phi} \sin(\frac{\theta}{2}) & -\cos(\frac{\theta}{2}) \\
          \end{pmatrix} \, ,
\ee
with the spherical coordinates $\theta$ and $\phi$, see Fig.~\ref{wig_rot}. In other words, the rotated states expressed in terms of the $z$ basis read
\begin{subequations}
\bal
\ket{\Psi^{+}} &= \cos\left(\frac{\theta}{2}\right) \ket{\Psi_z^{+}} + e^{i \phi} \sin\left(\frac{\theta}{2}\right) \ket{\Psi_z^{-}} \, , \\
\ket{\Psi^{-}} &= e^{-i \phi} \sin\left(\frac{\theta}{2}\right) \ket{\Psi_z^{+}} -\cos\left(\frac{\theta}{2}\right) \ket{\Psi_z^{-}} \, .
\eal
\end{subequations}
As a result, the transition amplitude valid for any spin quantization axis can be written as
\be
\label{amplitude_any_spin}
M_{s \rightarrow s'} = \sum_{i,j} \mathcal{D}_{j s'}^{*} M_{i \rightarrow j}^z \mathcal{D}_{i s} \, ,
\ee
with $M_{i \rightarrow f}^z$ being the transition amplitude when the spin quantization axis is chosen along the $z$-axis. Furthermore, the spin resolved differential ionization rate for an arbitrary spin quantization axis can be found via
\be
\label{rate_any_spin}
\frac{d W_{s \rightarrow s'}}{d^3 \vec{p}} = \frac{\omega}{\pi} \left( \sum_{i,j,k,l} \mathcal{D}_{j s'}^{*} \mathcal{D}_{l s'} M_{i \rightarrow j}^z M_{k \rightarrow l}^{z \,*} \mathcal{D}_{i s} \mathcal{D}_{k s}^{*} \right) \, .
\ee

\subsection{Dressed cc-SFA} \label{subsec:dressed_state}

Although, s-cc-SFA improves the results in comparison to the usual SFA via the Coulomb correction to the continuum electron wave function, it neglects the influence of the laser field on the bound state. Because of that, the electron spin dynamics in the bound state is completely neglected and the electron spin in the tunneling bound state is the same as in the initial state before the interaction with the laser field. In this case the spin effects are determined solely by the electron dynamics during the tunneling and during the motion in the continuum. Due to the evident asymmetry in the spin evolution in this picture (frozen spin in the bound state, rotating spin in the tunneling, and oscillating spin in the continuum) relatively large spin effects arise. On the other hand, it is well-known that the laser field can induce a significant spin dynamics in the bound state \cite{Hu_1999}. Moreover, the Zeeman-splitting of bound state levels can have an impact on the tunneling probabilities, in this way modifying the spin effects \cite{popov_2004u}. Therefore, it is important to take into account the laser field influence on the spin evolution in the bound state when calculating spin effects in the ionization process.

With this  motivation, we employ d-cc-SFA, see paper II, which is based on a specific partition of the Hamiltonian, in which the bound state dynamics in the laser field is accounted for. In  d-cc-SFA the total Hamiltonian is split up  as follows
\bal
H & = H_0^d + H_{int}^d
\, , \\
H_0^d & = c \vec{\alpha} \cdot \vec{p} + \beta c^2  + V(r) - (\vec{x} \cdot \vec{E})( \vec{\alpha}
\cdot \hat{\vec{k}}) \, , \\
H_{int}^d & = \vec{x} \cdot \vec{E} \,.
\eal
In this case the SFA transition amplitude reads
\be
\label{cc_d_SFA_m_matrix}
M_{s \rightarrow s'} = - i \int_{-\infty}^\infty dt \int d^3 x \, {\Psi_C^{s'}}^\dagger(\vec{x},t) \, (\vec{x} \cdot \vec{E}) \,
\Phi_0^s (\vec{x},t)\, ,
\ee
where the so-called dressed ground state satisfies the following equation
\be
\label{dressed_ground_state_equation}
i \partial_t \ket{\Phi_0^s}  = H_0^d \ket{\Phi_0^s} \, .
\ee
This equation includes the spin precession in the bound state induced by the laser field.

The Schr\"{o}dinger equation for the dressed bound state~(\ref{dressed_ground_state_equation}) cannot be solved analytically in an exact way and, therefore, several approximations are applied. Since the typical dimension of the atomic bound state is much smaller than the wavelength of the laser, we apply the dipole approximation for treating the dressed bound state dynamics, i.e., $\eta \rightarrow t$. Furthermore, we are concerned mostly by the electron spin precession in the bound state, consequently, we consider only transitions in the subspace of spin states and describe the dressed ground state with the following ansatz
\be
\label{ground_state_dsfa}
\ket{\Phi_0^s (t)}= \sum_{s'} C^{s s'} (t) \ket{\psi_0^{s'} (t)} \, ,
\ee
where $\ket{\psi_0^{s'} (t)}$ is the ground state wave function whose position representation is given by Eq.~(\ref{ground_state}).
According to the wave equation (\ref{dressed_ground_state_equation}), the coefficients $C^{s s'}(t)$ satisfy the following differential equation
\be
i \dot{C}^{s s''}(t) = \sum_{s'} C^{s s'}(t) \bra{\psi_0^{s''} 
(t)}H_1(t)\ket{\psi_0^{s'} (t)} \, ,
\ee
where $H_1(t) =  - \vec{E}(t) \cdot \vec{x} \, \vec{\alpha} \cdot \hat{\vec{k}} \,$. When the spin quantization direction is the $z$-axis,
the value of the matrix elements can be written as
\begin{subequations}
\label{matrix_element_z}
\bal
\bra{\psi_0^{+} (t)}H_1 (t) \ket{\psi_0^{+} (t)} &= \bra{\psi_0^{-} 
(t)}H_1 (t)
\ket{\psi_0^{-} (t)} = 0 \, ,\\
\bra{\psi_0^{+} (t)}H_1 (t) \ket{\psi_0^{-} (t)} &= \bra{\psi_0^{-} 
(t)}H_1 (t)
\ket{\psi_0^{+} (t)}^{*} =  i \frac{i E_y (t) - E_x (t)}{2c} \delta \, ,
\eal
\end{subequations}
where $\delta\equiv 1- 2 I_p/(3c^2)$. Thus, we obtain the following coupled differential equations
\begin{subequations}
  \label{ce}
\bal
\dot{C}^{+\pm} (t) &= C^{+\mp} (t) F^{\mp} (t) \,, \\
\dot{C}^{-\pm} (t) &= C^{-\mp} (t) F^{\mp} (t) \,, 
\eal
\end{subequations}
with
\be
F^{\pm}(t) = \frac{i E_y (t) \pm E_x (t)}{2c / \delta} \, .
\ee

It is useful to separate the bound state propagation in two stages; first, from the switching on the laser field up to the ionization moment, and  second, the under-the-barrier dynamics. The description of the first stage requires the exact solution of Eq. (\ref{ce}), which for linear and circular polarizations of the laser field are discussed in Sec.\ref{sec:bound_state}. For the second stage of the under-the-barrier dynamics during the imaginary time, a simple approximate solution  of the bound state wave function can be found exploiting the shortness of this time propagation. In fact, the time integration in Eq.~(\ref{cc_d_SFA_m_matrix}) will be performed by SPA and the matrix element (\ref{cc_d_SFA_m_matrix}) will be evaluated at the saddle point $\eta_s = \eta_r + i \eta_i$ (see Sec.\ref{mom_at_saddle}), where $\eta_r$ is the instant of ionization at which the spin state leaves the bound state, and $\eta_i$ is the duration of the imaginary time (the Keldysh time), during which the state evolves under-the-barrier, yet, it is small with respect to the laser period in the tunneling regime $|\omega \eta_i|\sim \gamma \ll 1$ [the tunneling regime of ionization is defined by the Keldysh parameter $\gamma=\sqrt{2I_p}\omega/E_0\ll 1$]. 

In order to investigate solely the under-the-barrier propagation, we first convert these differential equations to  Volterra integral equations of the second kind, which for the time propagation after $t=t_r$ can be written as
\begin{subequations}
\label{Volterra_0}
\bal
C^{+ \pm} (t) &= C^{+ \pm}_0 + C^{+ \mp}_0 \int_{t_r}^t ds \, F^{\mp}(s) \\
\nonumber & + \int_{t_r}^t ds \, C^{+ \pm}(s) F^{\mp}(s) \int_s^t d \tau \, F^{\pm}(\tau)  \, , \\
C^{- \pm} (t) &= C^{- \pm}_0 + C^{- \mp}_0 \int_{t_r}^t ds \, F^{\mp}(s) \\
\nonumber & + \int_{t_r}^t ds \, C^{- \pm}(s) F^{\mp}(s) \int_s^t d \tau \, F^{\pm}(\tau)  \, ,
\eal
\end{subequations}
where $C^{ss'}_0 \equiv C^{ss'} (t_r)$.
For calculation of the maximal tunneling probability, we choose $t_r = 0$, which corresponds to the electric field maximum, see Eq.~(\ref{e_field}). Then, introducing the dimensionless parameter $\varphi \equiv \omega t$, we arrive at the following integral equation
\be
\label{Volterra}
C^{s s'} (\varphi) = f^{s s'}(\varphi) + \frac{1}{4} \delta ^2 \xi ^2 \int_{0}^\varphi d u  K^{s s'} (\varphi,u) C^{s s'}(u)  \, ,
\ee
with the relativistic invariant field parameter $\xi \equiv E_0 / (c \omega)$. Here, the integral Kernels are given by
\bal
\label{KKK} K^{\pm +} (\varphi, u) & = {K^{\pm -}}^* (\varphi, u)  = \left[\cos (u)+i \zeta  \sin (u)\right] \\
\nonumber & \times \left( i \zeta  \left[\cos (u)-\cos (\varphi)\right]+\sin (u)-\sin (\varphi) \right) \, , 
\eal
and the functions $f^{s s'}(t)$ are 
\begin{subequations}
\label{FFF}
\bal
f^{+\pm}(\varphi) & = C^{+\pm}_0 +C^{+\mp}_0 \frac{\delta  \xi}{2}   \left[i \zeta  (\cos (\varphi)-1) \pm \sin (\varphi) \right] \, , \\
f^{-\pm}(\varphi) & = C^{-\pm}_0 +C^{-\mp}_0 \frac{\delta  \xi}{2}   \left[i \zeta  (\cos (\varphi)-1) \pm \sin (\varphi) \right] \, .
\eal
\end{subequations}

The formal solution to the integral equation~(\ref{Volterra}) can be given by the iterative method \cite{Polyanin}. Although in weak laser fields, $\xi \ll 1$, a few iterations would give an accurate result, for the strong fields, $\xi \gg 1$, generally, one has to deal with the infinite sum of iterations. Nevertheless, we consider a very short time propagation $\varphi= \omega \eta_i  \ll 1$. Moreover, we observe from Eq.~(\ref{Volterra})-(\ref{FFF}) that the effective time parameter scales as $\xi\delta \varphi$, which can be estimated as $ \xi \delta \gamma \sim \sqrt{I_p}/c \ll 1$ and it is small. Therefore, the solution of the integral equation~(\ref{Volterra}) for the considered small time propagation can be represented by the leading iteration term even for strong fields. The first iteration, which leads already to terms of the order of $(I_p/c^2)^{3/2}$, is given by the following expressions
\begin{subequations}
\label{solution_volterra}
\bal
C^{+\pm}(\varphi) &= C^{+\pm}_0 \pm\frac{C^{+\mp}_0}{2} \delta \xi \varphi - \frac{1}{8} \left[2 i C^{+\mp}_0 \delta \xi \zeta + C^{+\pm}_0 \delta^2 \xi^2 \right] \varphi^2 \\
\nonumber &  \mp\frac{1}{48}   \left[4 C^{+\mp}_0 \delta \xi -2 i C^{+\pm}_0 \delta^2 \xi^2 \zeta + C^{+\mp}_0 \delta ^3  \xi ^3  \right] \varphi^3   \, , \\
C^{-\pm}(\varphi) &= C^{-\pm}_0 \pm\frac{C^{-\mp}_0}{2} \delta \xi \varphi - \frac{1}{8} \left[2 i C^{-\mp}_0 \delta \xi \zeta + C^{-\pm}_0 \delta^2 \xi^2 \right] \varphi^2 \\
\nonumber &  \mp\frac{1}{48}   \left[4 C^{-\mp}_0 \delta \xi -2 i C^{-\pm}_0 \delta^2 \xi^2 \zeta + C^{-\mp}_0 \delta ^3  \xi ^3  \right] \varphi^3   \, . 
\eal
\end{subequations}
Furthermore, the coefficients~(\ref{solution_volterra}) can be represented as
\be
C^{s s' } (\varphi) = \Pi(\varphi, 0) C^{s s'}_0 \, , 
\ee
with the transformation matrix
\be
\Pi(\varphi,0) = \begin{pmatrix}
            1 - \frac{1}{8} \delta ^2 \xi ^2 \varphi^2  &  \frac{1}{2} \delta  \xi  \varphi -\frac{1}{4} i \delta  \zeta  \xi  \varphi^2 \\ 
		+ \frac{1}{24} i \delta ^2 \zeta  \xi ^2 \varphi^3  &  - \left( \frac{1}{48} \delta ^3 \xi ^3 +\frac{1}{12} \delta  \xi \right) \varphi^3 \\ 
		  & \\
            - \frac{1}{2} \delta  \xi  \varphi -\frac{1}{4} i \delta  \zeta  \xi  \varphi^2 & 1 - \frac{1}{8} \delta ^2 \xi ^2 \varphi^2   \\
	    + \left( \frac{1}{48} \delta ^3 \xi ^3 +\frac{1}{12} \delta  \xi \right) \varphi^3 &  - \frac{1}{24} i \delta ^2 \zeta  \xi ^2 \varphi^3 \\
          \end{pmatrix} \, .
\ee
In this way we separate the propagation of the spin states through the imaginary time axis from the states at the instant of ionization.

Then following the same procedure as in the case of s-cc-SFA, the transition amplitude in d-cc-SFA for the case when the spin quantization direction is the $z$-direction can be written as
\bal
\nonumber M_{s \rightarrow s'}^{z} &= \tilde{N}  \frac{\exp\left(i
\tilde{S}(\eta_s)\right)}{\left[\vec{q}(\eta_s) \cdot \vec{E}(\eta_s)\right]^{3/2}} 
\frac{|E(\eta_s)|^{I_p/c^2}}{\sqrt{\lambda}}  \\
\label{m_matrix_cc_d_SFA}& \times v_{s'}^{z \, \dagger} \left[1 + \frac{1}{2c \lambda} \vec{\alpha} \cdot
\vec{A}(\eta_s) (1 + \vec{\alpha}\cdot \hat{\vec{k}})
\right] \sum_{s''} C^{s s''}_0  U_{s''}^z (\eta_s) \, ,
\eal
where 
\bal
& U_{\pm}^z (\eta_s) = \left[ 1 - \frac{1}{8} \delta ^2 \xi ^2 (\omega \eta_s)^2 \pm \frac{1}{24} i \delta ^2 \zeta  \xi ^2 (\omega \eta_s)^3 \right] \tilde{u}_{\pm}^z(\eta_s)  \, \nonumber\\
\nonumber & \mp \left[ \frac{1}{2} \delta  \xi  (\omega \eta_s) \pm\frac{1}{4} i \delta  \zeta  \xi  (\omega \eta_s)^2 - \left( \frac{\delta ^3 \xi ^3 }{48} +\frac{\delta  \xi}{12}  \right) (\omega \eta_s)^3  \right] \tilde{u}_{\mp}^z(\eta_s) \, .
\eal

The transition amplitude as well as the differential ionization rate for any spin quantization axis can be defined via Eqs.~(\ref{amplitude_any_spin}) and (\ref{rate_any_spin}), respectively.

\section{Spin dynamics in the bound state } \label{sec:bound_state}

The spin resolved ionization amplitudes in d-cc-SFA, given by Eq.~(\ref{m_matrix_cc_d_SFA}), depend on the coefficients $C^{ss'}_0$, which describe the spin precession in the bound state when the quantization axis is along the laser propagation direction and are evaluated at the ionization time $t_r$. These coefficients are found from the solution of the system of differential equations~(\ref{ce}) with the following boundary conditions
\begin{subequations}
\label{initial_conditions}
\bal
\lim_{\xi \to 0} C^{\pm \pm} (t)&  =  1 \, , \\
\lim_{\xi \to 0} C^{\pm \mp} (t) & = 0 \, ,
\eal
\end{subequations}
describing the initial spin state of the atom when the laser field is switched off adiabatically. 

\subsection{Circular polarization} \label{bound_state_dynamics_circ}

In the case of circular polarization, the solution of Eq.~(\ref{ce}), which satisfy the boundary conditions~(\ref{initial_conditions}), is
\begin{subequations}
\label{circ_exact}
\bal
& C^{\pm\pm}_{circ} (t) = \frac{1+ \sqrt{1+\delta^2 \xi^2}}{\sqrt{\delta ^2 \xi ^2+\left(1+ \sqrt{1+\delta^2 \xi^2}\right)^2}} e^{\mp \frac{1}{2} i t \omega  \left(1-\sqrt{1 + \delta ^2 \xi ^2}\right)} \, , \\
& C^{\pm\mp}_{circ} (t) = \frac{i\delta\xi}{ \sqrt{\delta ^2 \xi ^2+\left(1+ \sqrt{1+\delta^2 \xi^2}\right)^2}} e^{\pm\frac{1}{2} i t \omega  \left(1 + \sqrt{1 + \delta ^2 \xi ^2}\right)} \, .
\eal
\end{subequations}
The maximum of the ionization probability is at the instant of ionization $t_r=0$, see Eq.~(\ref{e_field}), at which the coefficients yield
\begin{subequations}
\label{circ_at_instant}
\bal
C^{++}_{circ} (t_r=0) & = C^{--}_{circ} (t_r=0) = \frac{1+ \sqrt{1+\delta^2 \xi^2}}{\sqrt{\delta ^2 \xi ^2+(1+ \sqrt{1+\delta^2 \xi^2})^2}} \, , \\
C^{+-}_{circ} (t_r=0) & = C^{-+}_{circ} (t_r=0) = \frac{i\delta\xi}{ \sqrt{\delta ^2 \xi ^2+(1+ \sqrt{1+\delta^2 \xi^2})^2}} \, .
\eal
\end{subequations}
In the weak field limit, $\xi \ll 1$, one has the following asymptotics
\begin{subequations}
\label{circ_weak}
\bal
& C^{++} (t_r=0) = C^{--} (t_r=0) = 1 -\frac{1}{8} \xi^2 \delta^2 \, ,  \quad  \xi \ll 1 \, , \\
& C^{+-} (t_r=0) = C^{-+} (t_r=0) = \frac{i}{2} \delta \xi  - \frac{3i}{16} \delta^3 \xi^3 \, ,  \quad  \xi \ll 1 \, ,
\eal
\end{subequations}
while in the strong field regime, $\xi \gg 1$  
\begin{subequations}
\label{circ_strong}
\bal
C^{++}_{circ} (t_r=0) & = C^{--}_{circ} (t_r=0) = \frac{1}{\sqrt{2}} \, ,  \quad  \xi \gg 1 \, , \\
C^{+-}_{circ} (t_r=0) & = C^{-+}_{circ} (t_r=0) = \frac{i}{\sqrt{2}} \, ,  \quad  \xi \gg 1 \, .
\eal
\end{subequations}

We observe from the solution~(\ref{circ_exact}) that the spin oscillates over time with the frequency $\omega  \left(1 \pm \sqrt{1 + \delta ^2 \xi ^2}\right)/2$. While for weak fields it oscillates around the value at $t_r = 0$~(\ref{circ_weak}) with the laser's frequency $\omega$, for strong fields the spin state does full oscillation between the up and down states with a large frequency $ \delta \xi \omega / 2 $. As the coefficients $C_{circ}^{ss'}(t_r)$ and, accordingly, the spin resolved ionization probability oscillate with respect to the ionization time $t_r$, it is physically more appropriate to consider the time averaged ionization rate. The latter, in fact, corresponds to the  averaging over the photoelectron momentum,  because there is a mapping between the final momentum and the instant of ionization  via the saddle point equation. Choosing the interval of the time averaging $T$ much smaller than the laser period $T_0$, one can still relate the averaged ionization probability to the maximum of the momentum distribution of the tunneling electron.

In the weak field regime the averaging result coincides with the instantaneous one with an accuracy of the order of $T/T_0$ ($T/T_0\ll 1$). However, in the strong field regime the period of the spin oscillations are $\xi$ times smaller than the laser period. In this case we average the ionization probability over the time period of the spin oscillation $T=T_0/\xi$.

From Eq.~(\ref{m_matrix_cc_d_SFA}) one can see that the spin dependent part of the transition amplitude in the case of d-cc-SFA is determined by the following factor
\be
\label{spin_resolved_M}
\mathcal{S}_{s\rightarrow s'}^z \equiv  V_{s'}^{z \,\dagger} (\eta'_s) \sum_{s''} C^{s s''} (t_r)  U_{s''}^z (\eta'_s) \, ,
\ee
with 
\be
V_{s'}^{z \, \dagger} (\eta'_s) \equiv v_{s'}^{z\, \dagger} \left[1 + \frac{1}{2c \lambda} \vec{\alpha} \cdot
\vec{A}(\eta'_s) (1 + \vec{\alpha}\cdot \hat{\vec{k}}) \right] \, .
\ee
Here, we did the replacement $C^{s s''}_0 \rightarrow C^{s s''} (t_r)$ as well as $\eta'_s \rightarrow t_r + \eta_s$ in order to investigate effect of an arbitrary instant of ionization $t_r$. Then, the averaged differential ionization rate contains the factor
\bal
\label{averaged_diff_ionization_rate_circ}
\nonumber & \langle  | \mathcal{S}_{s \rightarrow s'}^{z\, \, circ} |^2 \rangle  = \langle  |\tilde{\mathcal{S}}_{+ \rightarrow s'}^z (\eta'_s) C^{s +}_{circ} (t_r) + \tilde{\mathcal{S}}_{- \rightarrow s'}^z (\eta'_s) C^{s -}_{circ} (t_r)|^2 \rangle  \, , \\
\nonumber & =   |\tilde{\mathcal{S}}_{+ \rightarrow s'}^z (\eta'_s)|^2 \langle  |C^{s +}_{circ} (t_r)|^2\rangle  +  |\tilde{\mathcal{S}}_{- \rightarrow s'}^z (\eta'_s)|^2 \langle |C^{s -}_{circ} (t_r)|^2\rangle    \\
\nonumber & + \tilde{\mathcal{S}}_{+ \rightarrow s'}^z (\eta'_s) \tilde{\mathcal{S}}_{- \rightarrow s'}^{z \, *} (\eta'_s)  \langle C^{s + }_{circ}(t_r) {C^{s -}_{circ}}^{*} (t_r)\rangle  \\
 & + \tilde{\mathcal{S}}_{- \rightarrow s'}^z (\eta'_s) \tilde{\mathcal{S}}_{+ \rightarrow s'}^{z\, *} (\eta'_s) \langle C^{s - }_{circ} (t_r) {C^{s +}_{circ}}^{*} (t_r) \rangle    \, , 
\eal 
with  $\tilde{\mathcal{S}}_{s \rightarrow s'}^z (\eta'_s) \equiv V_{s'}^{z \, \dagger} (\eta'_s) U_{s}^z (\eta'_s)$. Here, the slow oscillating functions $\tilde{\mathcal{S}}_{s \rightarrow s'}^z (\eta_s)$ with respect to the averaging time $T$ are taken out from the averaging. From Eq.~(\ref{circ_exact}) one can see that the mean values $|C^{s \pm}_{circ} (t_r)|^2 $ are time independent, and the frequency of the oscillations of $C^{s \pm }_{circ} (t_r) {C^{s \mp}_{circ}}^{*}(t_r)$ is $\omega$. Therefore, the averaged probability over the time $T\ll T_0$ will coincide with the instantaneous value
\be
 \langle  | \mathcal{S}_{s \rightarrow s'}^{z\, \, circ} |^2 \rangle  = | \mathcal{S}_{s \rightarrow s'}^{z\, \, circ} (t_r=0)|^2 \, .
\ee

However, this particular choice of the spin quantization axis along the laser propagation direction is very special in the circular polarization case. When the quantization axis is arbitrary, the probability is a linear combination of $C^{s_1 s_2}_{circ} (t_r) {C^{s_3 s_4}_{circ}}^{*} (t_r)$ with arbitrary values of $s_i$ because 
\bal
\nonumber 
\langle  | \mathcal{S}_{s \rightarrow s'}^{circ} |^2 \rangle  &= \langle  
\sum_{i,j,k,l} \mathcal{D}_{j s'}^{*} \mathcal{D}_{l s'} \mathcal{D}_{i s} \mathcal{D}_{k s}^{*} \\ 
\nonumber & \times \left( \tilde{\mathcal{S}}_{+ \rightarrow j}^z (\eta'_s) C^{i +}_{circ} (t_r) + \tilde{\mathcal{S}}_{- \rightarrow j}^z (\eta'_s) C^{i -}_{circ} (t_r) \right) \\
\nonumber & \times \left( \tilde{\mathcal{S}}_{+ \rightarrow l}^z (\eta'_s) C^{k +}_{circ} (t_r) + \tilde{\mathcal{S}}_{- \rightarrow l}^z (\eta'_s) C^{k -}_{circ} (t_r) \right)^{*} \rangle  \, ,
\eal
which, for instance, contains a term
\be
\tilde{\mathcal{S}}_{- \rightarrow j}^z(\eta_s) \tilde{\mathcal{S}}_{+ \rightarrow l}^{z \, *}(\eta_s) \langle  C^{+ -}_{circ} (t_r) {C^{- +}_{circ}}^{*} (t_r) \rangle  \, .
\ee
The term $C^{+ -}_{circ} (t_r) {C^{- +}_{circ}}^{*} (t_r)$ oscillates with the frequency $\xi \delta \omega$ for strong fields, and its mean value  is vanishing  $\langle C^{+ -}_{circ} (t_r) {C^{- +}_{circ}}^{*} (t_r) \rangle = 0$ when the 
averaging period fulfills the condition
\be
\label{period_condition}
\frac{1}{\delta \xi \omega}\ll T \ll T_0 \, .
\ee
Note that that the instantaneous value of this term at $t_r=0$ is $1/2$ in the strong field limit, $\xi \gg 1$. Therefore, the averaged probability will differ from the instantaneous value in those cases when the direction of quantization axis is chosen other than in the laser propagation direction.

\subsection{Linear polarization}

In the case of  linear polarization of the laser field, the solution of Eq.~(\ref{ce}), with the boundary conditions according to Eq.~(\ref{initial_conditions}), is 
\begin{subequations}
\label{lin_exact}
\bal
C^{\pm \pm}_{lin} (t) & =  \cos \left[\frac{1}{2} \delta  \xi  \sin (\omega t )\right] \, ,\\
C^{\pm \mp}_{lin} (t) &= \mp \sin \left[\frac{1}{2} \delta  \xi  \sin (\omega t )\right] \, .
\eal
\end{subequations}
The spin states at the instant of ionization, $t_r = 0 $ are
\begin{subequations}
\label{lin_at_instant}
\bal
C^{\pm \pm}_{lin} (t_r=0) & = 1 \, ,\\
C^{\pm \mp}_{lin} (t_r=0) &= 0 \, .
\eal
\end{subequations}

Similar to the circular polarization case, in weak fields  $\xi \ll 1$, the spin states oscillate over time around the value at the instant of ionization $t_r=0$  with the laser's frequency,  
\begin{subequations}
\label{lin_abs2_waek}
\bal
C^{\pm \pm}_{lin} (t) &=  1 - \frac{1}{2}\left( \delta  \xi  \sin (\omega t ) /2 \right)^2  \, ,  \quad  \xi \ll 1 \, , \\
C^{\pm \mp}_{lin} (t) &=  \mp \frac{1}{2} \delta  \xi  \sin (\omega t )  \, ,  \quad  \xi \ll 1 \, ,
\eal
\end{subequations}
Consequently, in weak fields  $\xi \ll 1$ the time averaged probability, carried out over the time region given by Eq.~(\ref{period_condition}), coincides with the instantaneous one.

In strong fields  $\xi \gg 1$,  the spin oscillates between the up and down states with the frequency $\delta \xi \omega / 2 $. The mean value of the differential ionization rate for an arbitrary spin quantization axis in the linear polarization case  includes the following three factors; $\langle \cos \left[\frac{1}{2} \delta  \xi  \sin (\omega t_r )\right] \sin \left[\frac{1}{2} \delta  \xi  \sin (\omega t_r )\right]\rangle =0$, $\langle \cos^2 \left[\frac{1}{2} \delta  \xi  \sin (\omega t_r )\right]\rangle =1/2$, and $\langle \sin^2 \left[\frac{1}{2} \delta  \xi  \sin (\omega t_r )\right]\rangle =1/2$, when the averaging is carried out over the time region (\ref{period_condition}). Accordingly, when the quantization axis is along the $z$-direction, the averaged spin resolved ionization probability in the strong field limit is determined by the factor  
\bal
 \langle | \mathcal{S}_{s \rightarrow s'}^{z\, \, lin} |^2 \rangle  & = \frac{1}{2} \left( |\tilde{\mathcal{S}}_{+ \rightarrow s'}^z (\eta_s)|^2  + 
 |\tilde{\mathcal{S}}_{- \rightarrow s'}^z (\eta_s)|^2 \right) \, , \quad \xi \gg 1 \, ,
\eal
which does not coincide with the instantaneous value of the transition probability.
The rate for an arbitrary quantization axis can be calculated in a similar way. 

As a summary, the spin resolved differential ionization rate corresponding to the maximum of the photoelectron momentum distribution is well defined in the case of weak fields and corresponds to the instantaneous ionization rate at the peak of the laser field. In strong laser fields, the spin resolved  ionization rate is highly oscillating with respect to the ionization time, therefore, it is physically relevant to average the rate  over a period fulfilling the condition~(\ref{period_condition}).

\section{Final momentum distribution of the tunnel-ionized electron} \label{mom_at_saddle}

We are concerned with the spin resolved ionization probabilities corresponding to the maximum of the momentum distribution of  photoelectrons. In this section we derive the momentum distribution of the directly ionized electrons, taking into account the relativistic corrections during the under-the-barrier motion ($\sim I_p/c^2$) as well as the nonadiabatic corrections ($\sim \gamma^2$). The momentum corresponding to the maximum of this distribution will be used in the next section for the evaluation of the spin asymmetries as well as for the spin flip. 

In SFA the dominating part of the ionization probability is given by the tunneling exponent
\be
\label{probability_s}
W \sim \left|\exp\left( i \tilde{S} \left(\vec{p},\eta_s(\vec{p})\right) \right) \right|^2 \, ,
\ee
where the exponent depends on the final momentum $\vec{p}$ as well as the saddle point $\eta_s = \eta_r + i \eta_i$. Furthermore, via the saddle point equation $\vec{q}(\eta_s)^2 = - \kappa^2$, the momentum and the saddle point are connected to each other so that the exponent is a function of the final momentum. In other words, each saddle pints correspond to different momenta for the tunnel-ionized electron. In order to find the momentum that maximizes the ionization probability, we first note that in the quasistatic tunnel-ionization regime ($\gamma\ll 1$) the tunneling probability is maximal when the electric field reaches the maximum at the real part of the saddle time, i.e., when
\be
\label{max_tunneling_condition}
| \vec{E}(\eta_r) | = E_0 \, .
\ee
Here $\eta_r$ corresponds to the instant of ionization in the quasiclassical description of tunneling such that the electron leaves the bound state and starts the continuum motion. Then, taking into account Eqs.~(\ref{e_field}) and (\ref{b_field}) we find $\eta_r = 0$, and 
\bal
\label{e_field_at_exit}
\vec{E}(\eta_r) &= - E_0 \hat{\vec{x}} \, , \\
\label{b_field_at_exit}
\vec{B}(\eta_r) &= - E_0 \hat{\vec{y}} \, , \\
\label{a_field_at_exit}
\vec{A}(\eta_r) &= - \frac{E_0}{\omega} \zeta \hat{\vec{y}} \, .
\eal
Note that for a circularly polarized field, at any time the condition~(\ref{max_tunneling_condition}) is fulfilled. Nevertheless, in order to generalize the result to an arbitrary polarization, we set $\eta_r = 0$. This condition further implies that the most probable tunneling is along the direction that points the maximal electric  field.  Within this conclusion, the saddle point equation  (\ref{saddle_point_equation}) reads
\bal
\nonumber & \kappa^2 + c^2\left(1-\lambda - I_p/c^2 \right)^2 + \left(p_y - \frac{E_0 \zeta \cosh(\omega \eta_i)}{\omega} 
\right)^2  \\
\label{saddle_point_equation_1} & + \left(p_x + i \frac{E_0 \sinh(\omega \eta_i)}{\omega}\right)^2 = 0 \, .
\eal
For a real $\eta_i$, Eq.~(\ref{saddle_point_equation_1}) can only be fulfilled if 
\be
\label{ass_px}
p_x = 0 \, ,
\ee
i.e., the final momentum along the tunneling direction of the tunnel-ionized electron should vanish. Correspondingly, the contracted action~(\ref{contracted_action}) can be written as
\bal
\nonumber \tilde{S} (\eta_s) &= \frac{E_0^2 (\zeta^2-1)\left[\sin(2 \omega \eta_s)-2 \omega \eta_s \cos(2 \omega \eta_s) \right]}{8 \lambda \omega^3} \\
 & - \frac{8 E_0 p_y \zeta \omega \left[ \sin(\omega \eta_s)-\omega
\eta_s \cos(\omega \eta_s) \right]}{8 \lambda \omega^3} \, .
\label{contracted_action_1}
\eal
Further, it is more convenient to maximize the tunneling probability not by the final momentum $\vec{p} \equiv \vec{p} (\eta_f)$ but by the momentum $\vec{p} (\eta_r)$ at the tunnel exit, assuming the electron propagation from the tunnel exit to the detector is governed by the laser field only. The latter is defined via the following relation,
\be
\label{eom_vector}
\vec{p} = \vec{p} (\eta_r) - \vec{A} (\eta_r)  - \frac{\hat{\vec{k}}}{\lambda c} 
 \left(
\vec{p} (\eta_r)  - \frac{\vec{A} (\eta_r)}{2}\right) \cdot \vec{A} (\eta_r)  \, ,
\ee
where we have used Eq.~(\ref{eom}) in Appendix~\ref{ceom}, and the fact that $\vec{A}(\eta_f) = 0$. Taking into account Eqs.~(\ref{a_field_at_exit}) and (\ref{ass_px}),  the relation (\ref{eom_vector}) yields
\begin{subequations}
\label{ass_p}
\bal
p_x & = p_x (\eta_r)=0 \, , \\
p_y & = p_y (\eta_r) + \frac{\zeta \, E_0}{\omega} \, , \\
p_z & = p_z (\eta_r) + \frac{1}{\lambda \, c} \frac{\zeta \, E_0}{\omega} \left(p_y (\eta_r) +
\frac{\zeta \, E_0}{2 \omega}\right) \, .
\eal
\end{subequations}
We observe that the initial momentum along the tunneling direction at the tunnel exit, $p_x (\eta_r)$, vanishes, which agrees with the simpleman model as $\eta_r$ is the turning point, i.e., the tunnel exit point, where the classical particle has a vanishing velocity.

Next we express the constant of motion $\lambda$ via $(p_y (\eta_r),p_z (\eta_r))$ as
\bal
\nonumber \lambda &= \frac{1}{c} \left( \sqrt{c^2 + (p_y (\eta_r)^2 +p_z (\eta_r)^2)}  - p_z (\eta_r) \right) \, ,
\\
& \approx  1 - \frac{p_z (\eta_r)}{c} \, ,
\eal
where we neglect ${p_y} (\eta_r)^2/c^2$ and ${p_z} (\eta_r)^2/c^2$ which are of the orders of $(E_0/E_a)(2I_p/c^2) $ [the latter follows from the fact that the width of the transverse momentum distribution at tunnel-ionization is $\sqrt{E_0}/(2I_p)^{1/4}$ \cite{Popov_2004}. Note that in the tunneling regime $E_0/E_a \le 1/10$, with the atomic field strength $E_a = (2I_p)^{3/2}$]. Using the $\lambda$ value, the contracted action~(\ref{contracted_action_1}) reads
\be
\tilde{S}(\eta_s) = -\frac{E_0\left[E_0 + \zeta \omega p_y (\eta_r)\right]}{3[1 -
p_z (\eta_r)/c]} \eta_s^3 \, ,
\ee
where we expand the action~(\ref{contracted_action_1}) up to nonvanishing  order in $\eta_s$ taking into account that $| \omega \eta_s |\sim \gamma \ll 1$. At the same order, the saddle point is calculated as
\be
\eta_s = i \sqrt{\frac{p_y (\eta_r)^2 + p_z (\eta_r)^2 + 2 I_p [1 - p_z (\eta_r)/c]}{E_0 [E_0 + \zeta p_y (\eta_r)\omega]}} \, .
\ee
\begin{figure}
  \centering
  \includegraphics[width=0.8\linewidth]{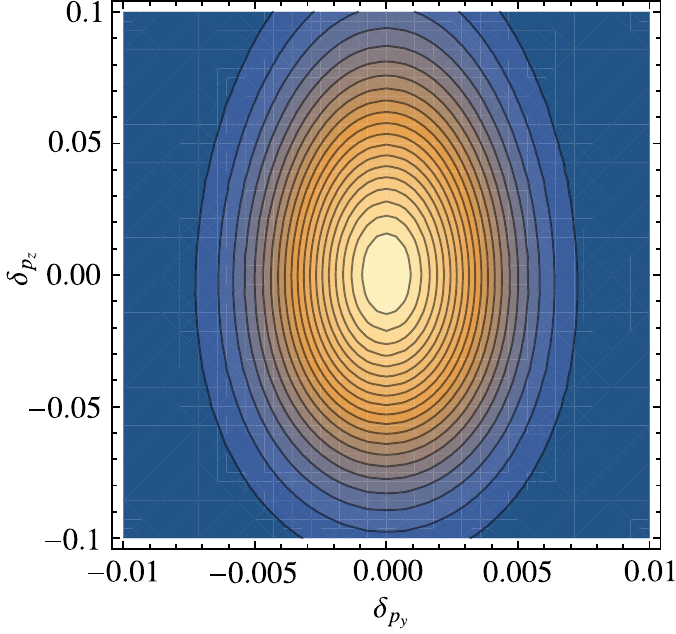}
  \caption{(Color online) Ionization probability calculated numerically via Eq. (\ref{probability_s}) versus the final transverse momentum. The final momentum is defined as $\vec{p}_{f\perp} \equiv \vec{p}_\perp (1 + \delta_{\vec{p}_\perp} ) $, with $\vec{p}_\perp$ given by Eq.~(\ref{max_mom_y}), and Eq.~(\ref{max_mom_z}) and $\delta_{\vec{p}_\perp}$ being the dimensionless deviation. The applied parameters are $\kappa = 50$, $\omega=1$, $\zeta=0.5$, $E_0 / E_a = 1 /30$. Note that we set $\omega=1$ instead of $\omega = 0.05$, which is a typical parameter for tunnel-ionization, in order to confirm that the derived analytical results are valid in a nonadiabatic regime as well.}
  \label{maximum_mom}
\end{figure}
Finally, we can identify the momentum which maximizes the tunneling probability via the condition
\be
\frac{\del \tilde{S}}{\del \vec{p}_\perp (\eta_r)} = 0 \, ,
\label{max_mom_condition}
\ee
with  $\vec{p}_\perp (\eta_r) = (p_y(\eta_r), p_z(\eta_r))$. The condition yields the following two equations
\begin{subequations}
\label{p_yz}
\bal
\label{p_y}
&  p_y (\eta_r)^2 - 2 p_z (\eta_r)^2 - I_p + c p_z (\eta_r) \left(3 + I_p/c^2  \right)=0  \, ,\\
\label{p_z}
& 6 E_0 p_y (\eta_r) + \zeta \omega \left\{5 p_y (\eta_r)^2 - p_z (\eta_r)^2 -  2I_p \left[1 -
\frac{p_z (\eta_r)}{c} \right] \right\}=0 \, .
\eal 
\end{subequations}
Up to the leading order in $I_p/c^2$, the solution of Eq.~(\ref{p_yz}) can be given by
\begin{subequations}
\label{initial_mom}
\bal
\label{initial_mom_y}
p_z (\eta_r) & = \frac{I_p}{3 c} \, , \\
\label{initial_mom_z}
p_y (\eta_r) & = \frac{\zeta \omega I_p }{3 E_0} \, .
\eal 
\end{subequations}

As a summary, the final momentum yielding the maximal tunneling probability, with
Eq.~(\ref{ass_p}), can be written as
\begin{subequations}
\label{max_mom_dist}
\bal
\label{max_mom_x}
 p_x & = 0 \, , \\
\label{max_mom_y}
 p_y & = \frac{\zeta \, E_0}{\omega} \left(1 + \frac{\gamma^2}{6} \right) \, , \\
\label{max_mom_z} 
p_z & = \frac{I_p}{3 c} + \frac{p_y^2}{2c}\left(1+ \frac{I_p}{3c^2}\right) \, ,
\eal
\end{subequations}
the validity of which is illustrated in  Fig.~\ref{maximum_mom}, where the final momentum distribution is calculated numerically via Eq. (\ref{probability_s}). We emphasize that the momentum distribution~(\ref{max_mom_dist}) is valid also for an arbitrarily elliptical polarization with an ellipticity $0 \le \zeta \le 1$.

Furthermore, by using the final momentum~(\ref{max_mom_dist}), we can write down the saddle point, which maximizes the tunneling probability,
as
\be
\label{saddle_point}
\eta_s = i \frac{\sqrt{2 I_p}}{E_0} \left[1 -\frac{5 I_p}{36 c^2} + \frac{\gamma ^2}{18} (\zeta
^2-3) \right] \, ,
\ee
where  $I_p/c^2$-expansion is applied. Accordingly, the maximal probability is derived from Eq.~(\ref{probability_s}) can be found as
\be
\label{probability_exponent}
W \sim \left|\exp\left( i \tilde{S} (\eta_s) \right)\right|^2 \approx \exp\left( - \frac{2 E_a}{3 E_0} 
\left[1 - \frac{I_p}{12 c^2}  + \frac{\gamma^2}{30} (\zeta^2 - 3)\right] \right) \, .
\ee

Here we should note that the initial transverse momentum given by Eq.~(\ref{initial_mom})  arises during the classically forbidden under-the-barrier dynamics in tunnel-ionization. The momentum along the laser propagation direction $I_p/ (3 c)$ is a relativistic signature of the process \cite{Yakaboylu_2013_rt}, whereas the momentum along the direction of the laser's magnetic field at the instant of ionization $\zeta E_0 \gamma^2 /(6 \omega)$, which depends on the laser polarization, is induced due to the nonadiabaticity of the ionization and it is significant at large Keldysh parameters $\gamma$  \cite{Smirnova_2011,Klaiber_2014}. In addition to the SFA prediction, there exist also an initial nonvanishing momentum along the tunneling direction for the most probable trajectory, if one goes beyond the quasiclassical description of tunneling and defines a tunneling time delay~\cite{Yakaboylu_2014_td}.

In the following sections, we discuss the spin dynamics in the tunneling regime, $\gamma \ll 1$, consequently, we will omit the $\gamma^2$ terms  in Eqs.~(\ref{max_mom_dist}) - (\ref{probability_exponent}) in the corresponding calculations.

\section{Spin asymmetries and spin flip} \label{sec:spin_asymmetries}

In order to investigate the spin dynamics we consider the following physically relevant choices of the spin quantization axis; along the laser propagation direction, along the direction of the laser electric field, and along the laser magnetic field at the instant of ionization.

Concerning the spin asymmetry during ionization, one may ask two independent questions: 
\begin{enumerate}
\item Does the ionization rate depend on the initial spin state of the bound electron? 
\item Will the electron be polarized after tunnel-ionization from an unpolarized target?
\end{enumerate}
Consequently, we can define two spin asymmetry parameters, firstly, the tunneling asymmetry parameter
\be
\label{asf}
\mathcal{A}_{t} = \frac{W_{+ \rightarrow +} + W_{+ \rightarrow -} - W_{- \rightarrow +} - W_{-
\rightarrow -}}{W_T} \, ,
\ee
which measures the asymmetry between the tunneling rates for initially polarized states. Here 
\be
W_{s \rightarrow s'}\equiv \left. \frac{d W_{s \rightarrow s'}}{d^3 \vec{p}}\right|_{\vec{p} = \vec{p}_M} \, ,
\ee
is the differential ionization rate at the momentum value $\vec{p} =\vec{p}_M$ corresponding to the maximal tunneling rate, see Eq.(\ref{max_mom_dist}), and 
\be
W_T= \frac{W_{+ \rightarrow +} + W_{+ \rightarrow -} + W_{- \rightarrow +} + W_{-
\rightarrow -}}{2}  \label{W_T}
  \, ,
\ee
is the total ionization rate averaged by the initial and summed over the final spin states.

Secondly, one may define the spin polarization asymmetry parameter as
\be
\label{ae}
\mathcal{A}_{p} = \frac{W_{+ \rightarrow +} + W_{- \rightarrow +} - W_{+ \rightarrow -} - W_{-
\rightarrow -}}{W_T} \, ,
\ee
which is a measure of the electron final polarization in the case of initially unpolarized states~\cite{Faisal_2004}.

In addition to the asymmetries we will provide also the spin flip rate 
\be
\mathcal{F}_\pm = \frac{W_{\pm \rightarrow \mp}}{W_T} \, .
\ee
The three independent parameters $\mathcal{A}_{t}$, $\mathcal{A}_{p}$ and $\mathcal{F}_+$ fully describe the spin transitions, taking into account the normalization condition $(1/2)\sum_{s,s'} W_{s \rightarrow s'} / W_T = 1$.

To sum up, we have introduced the following two sets of parameters: the laser polarization parameter $\zeta$, and the spin quantization angles $\theta, \phi$ in order to investigate analytical results for the spin effects. The latter are described by three independent parameters; the tunneling and polarization asymmetry parameters, and the spin flip, for which we discuss the weak field as well as the strong field limits in the following section.

\subsection{The s-cc-SFA prediction}

Let us first consider predictions of s-cc-SFA, which neglects the spin dynamics in the bound state. We underline that due the latter the results of s-cc-SFA are applicable for an arbitrary elliptical polarization $\zeta$.

\subsubsection{The  spin quantization axis is parallel to the laser propagation direction}

When the spin quantization axis is along the laser propagation direction~$\hat{\vec{z}}$, the angles determining the quantization axis are $\theta =0$ and $\phi=0$. In this case we obtain for the spin flip relative probability corresponding to the maximum of the momentum distribution
\be
\label{sf_s_SFA_k_direc}
\mathcal{F}_\pm^{z\,\,(s)} =  \frac{\zeta \xi (\zeta \xi/2 \mp  \rho)/2}{1 + \zeta^2 \xi^2/4}  + \mathcal{O}(\rho^2) \, ,
\ee
in the leading order of $\rho\equiv \sqrt{2I_p/c^2}$. On the one hand, the spin flip vanishes for a linearly polarized filed for the maximal tunneling probability. On the other hand, for a nonvanishing laser's polarization ($\zeta \ne 0$), the spin flip is negligible in the weak field regime $\xi\ll 1$,  $\mathcal{F}_\pm^{z\,\, (s)} \rightarrow 0$, whereas s-cc-SFA predicts almost complete spin flip for strong fields $\xi\gg 1$, $\mathcal{F}_\pm^{z\,\, (s)} \rightarrow 1$.

The tunneling asymmetry parameter and the spin polarization asymmetry parameter can be calculated up to the nonvanishing order of $\rho$ as 
\bal
\label{at_s_SFA_k_direc}
\mathcal{A}_{t}^{z\,\, (s)} &= - \frac{\zeta}{2 \xi}\rho^{3}  \, , \\
\label{ap_s_SFA_k_direc}
\mathcal{A}_{p}^{z\,\, (s)} & =  \frac{2 \zeta  \xi}{1+\zeta ^2\xi ^2/4}\rho \, .
\eal
The s-cc-SFA calculations predict a rather large spin polarization asymmetry because the parameter $\mathcal{A}_{p}^{z\,\, (s)}$ scales linearly with the parameter $\rho$, while the tunneling asymmetry parameter is much smaller $\sim \rho^3$. Nevertheless, the spin asymmetries disappear for the linear polarization case.

We note that on the one hand the relation $\rho^{3}/\xi =\rho^{2}\gamma$ indicates that there is no singularity in Eq.~(\ref{at_s_SFA_k_direc}) in the weak field regime ($\xi \ll 1$), where $\rho \ll \gamma\ll 1$. On the other hand, the relation $\rho^{3}/\xi = \omega/c^2 (E_a/E_0)$ indicates that Eq.~(\ref{at_s_SFA_k_direc}) vanishes for low frequencies. When neglecting the terms of the order of $\xi^2$  in the limit $\xi \ll 1$, one has also to neglect the terms of the order of $\rho^2$,  because they are much smaller than the former by the factor of $\gamma^2\ll 1$.

\subsubsection{The spin quantization axis is perpendicular to the laser propagation
direction}

In the case when the spin quantization axis  is perpendicular to the laser propagation direction, one can align it along the direction of the electric field or the magnetic field at the instant of ionization $\eta_r$, and for the sake of convenience we will consider the  directions opposite to the fields.  

When the  spin quantization axis is along the direction of $-\hat{\vec{E}}(\eta_r=0)$ ($\theta = \pi/2, \phi = 0$, see Fig.~\ref{wig_rot}), cf. Eq.~(\ref{e_field_at_exit}), the flip probability is 
\bal
\label{sf_s_SFA_e_direc}
\mathcal{F}_\pm^{x\,\, (s)} = \frac{\rho^2}{4} \, ,
\eal
which is tiny for nonrelativistic (low charge) ions. Furthermore, the asymmetries are vanishing
\bal
\label{atp_s_SFA_e_direc}
\mathcal{A}_{t}^{x\,\, (s)} =\mathcal{A}_{p}^{x\,\, (s)} =0
 \, .
\eal
The results are independent from both the polarization and intensity parameters at the maximum of the tunneling probability, $t_r=0$.

When the quantization axis is along the direction of $-\hat{\vec{B}}(\eta_r=0)$ ($\theta = \pi/2, \phi = \pi/2$), cf. Eq.~(\ref{b_field_at_exit}), the spin flip is 
\be
\label{sf_s_SFA_b_direc}
\mathcal{F}_\pm^{y\,\, (s)} =  \frac{(\zeta^2 \xi^2/4) (1 \pm \rho)}{1 + \zeta^2 \xi^2/4}  + \mathcal{O}(\rho^2) \, .
\ee
While it is insignificant for  weak fields, a complete spin flip occures in the strong field regime when $\zeta \ne 0$, which is similar to Eq.~(\ref{sf_s_SFA_k_direc}). On the other side, the spin asymmetries can be written as
\bal
\label{at_s_SFA_b_direc}
\mathcal{A}_{t}^{y\,\, (s)} &= 2\rho   \, , \\
\label{ap_s_SFA_b_direc}
\mathcal{A}_{p}^{y\,\, (s)} & =  2 \rho \frac{ 1- \zeta ^2 \xi
^2/4}{1+\zeta ^2 \xi ^2/4} \, .
\eal
The spin polarization asymmetry parameter as well as the tunneling asymmetry parameter scale as $\rho$. The latter is also independent from the polarization as well as the intensity parameter of the laser field.

\begin{table}
\centering
\begin{tabular}{l|l|l|l|l|l|l|l|l|}
 & \multicolumn{6}{l|}{Spin effects for strong fields $\displaystyle \xi \gg 1$}    \\ \hline
 & \multicolumn{2}{l|}{$\displaystyle  \hat{\vec{k}}$} & \multicolumn{2}{l|}{$\displaystyle  -\hat{\vec{E}}(\eta_r=0)$} &\multicolumn{2}{l|}{$\displaystyle  -\hat{\vec{B}}(\eta_r=0)$} \\ \hline
 &s-SFA &d-SFA &s-SFA &d-SFA &s-SFA &d-SFA \\ \hline
$\displaystyle \mathcal{F}_\pm$ & $\displaystyle \zeta $ & $\displaystyle 1/2 \pm \zeta \rho^3/4  $ & $\displaystyle \rho^2/4$ & $\displaystyle 1/2 $ & $\displaystyle  \zeta (1 \pm \rho) $ & $\displaystyle \zeta \left( 1/2 \pm  \rho^3/4\right) $  \\ \hline
$\displaystyle \mathcal{A}_{t}$ & $\displaystyle 0 $ & $\displaystyle \zeta \rho^3  $ & $\displaystyle 0$ & $\displaystyle 0 $ & $\displaystyle 2 \rho$ & $\displaystyle  (1- \zeta) \rho^3 $  \\ \hline
$\displaystyle \mathcal{A}_{p}$ & $\displaystyle 0$ & $\displaystyle 0  $ & $\displaystyle 0$ & $\displaystyle 0 $ & $\displaystyle  2 \rho (1-2 \zeta)$ & $\displaystyle \rho^3 (1 - 2 \zeta)  $  \\ \hline
\end{tabular}
\caption{Spin flip and asymmetries in strong field regime $\xi\gg 1$ at the instant of ionization associated to the maximal tunneling probability in the leading order of the parameter $1/\xi$. Comparison of the s-cc-SFA and d-cc-SFA results for different spin quantization axes at the leading order in $\rho \equiv \sqrt{2I_p}/c$; $\zeta =0$ is for linear  and $\zeta=1$ is for circular polarization of the laser field.}
\label{strong_field_table}
\end{table}

\subsection{The d-cc-SFA prediction}

In the weak field regime $\xi\ll 1$, as the asymmetries as well as the spin flip are negligible (smaller by an order of magnitude than $\rho^3$) for any choice of the quantization axes, in what follows, we discuss the predictions of the strong field limit of d-cc-SFA. Note that in contrast to the s-cc-SFA case, the polarization parameter is $\zeta = \{0,1\}$ for the d-cc-SFA case.

\subsubsection{The  spin quantization axis is parallel to the laser propagation direction}

In this case the spin flip in the strong field regime ($\xi\gg 1$) in the leading order of $\rho$ equals to
\be
\label{sf_d_SFA_k_direc}
\mathcal{F}_\pm^{z\, \, (d)}  = \frac{1}{2} \pm \zeta \frac{\rho^3}{4} \, ,
\ee
which is due to the spin fast oscillation in the bound state (see Sec.~\ref{simpleman_DSFA} for an intuitive description), and it is different than the prediction of the s-cc-SFA, whose strong field limit yields $\mathcal{F}_\pm^{z\,\, (s)} \rightarrow \zeta$ with $\zeta=\{0,1\}$. Furthermore, there is a correction at the order of $\rho^3$ for the circular polarization case, see Fig.~\ref{fig:SF_single}. The corresponding asymmetries  are
\bal
\label{at_d_SFA_k_direc}
& \mathcal{A}_{t}^{z\,\, (d)}  = \zeta \rho^3 + \mathcal{O}(\xi^{-1}) \, , \\
\label{ap_d_SFA_k_direc}
& \mathcal{A}_{p}^{z\,\, (d)} =   \mathcal{O}(\xi^{-1}) \, .
\eal
Firstly, the polarization asymmetry parameter vanishes for both of linear and circular polarizations. The tunneling asymmetry parameter also disappears in the linear polarization case, where s-cc-SFA and d-cc-SFA agree with each other for strong fields. However, the d-cc-SFA result for the tunneling asymmetry parameter differs from the s-cc-SFA one in the case of circular polarization, as it increases with increasing charge of the ion.

\subsubsection{The spin quantization axis is perpendicular to the laser propagation direction}

First of all, when we align the spin quantization axis along the $-\hat{\vec{E}}(\eta_r=0)$-direction, the spin flip can be calculated as
\be
\label{sf_d_SFA_e_direc}
\mathcal{F}_\pm^{x\, \, (d)}  =  \frac{1}{2} \, ,
\ee
which is again a consequence of the bound state dynamics, cf. Eq.~(\ref{sf_s_SFA_e_direc}). However, as in the case of s-cc-SFA, the asymmetries vanish
\be
\label{atp_d_SFA_e_direc}
\mathcal{A}_{t}^{x\,\,  (d)} = \mathcal{A}_{p}^{x \, \, (d)} = \mathcal{O}(\xi^{-1}) \, .  
\ee

In the case where the spin quantization axis is the $-\hat{\vec{B}}(\eta_r=0)$-direction, the spin flip is
\be
\label{sf_d_SFA_b_direc}
\mathcal{F}_\pm^{y\, \, (d)}  = \frac{\zeta}{2} \left( 1 \pm  \frac{\rho^3}{2}\right) + \mathcal{O}(\xi^{-1}) \, .
\ee
We first notice that  the spin flip vanishes for the linear polarization case. For a circularly polarized field the spin flip is the same as in the case of the quantization axis  along the $\hat{\vec{k}}$-direction.

The asymmetries in the d-cc-SFA case are
\bal
\label{at_d_SFA_b_direc}
\mathcal{A}_{t}^{y\,\,  (d)} & = (1- \zeta) \rho^3 + \mathcal{O}(\xi^{-1}) \,  , \\
\label{ap_d_SFA_b_direc}
\mathcal{A}_{p}^{y\, \, (d)} & = \rho^3 - 2 \zeta\rho^3 \, .
\eal
In the previous configuration, Eqs.~(\ref{at_d_SFA_k_direc}) and (\ref{ap_d_SFA_k_direc}) the both asymmetries vanish for a linearly polarized field, however in this case, the both have the same nonvanishing value. Moreover, we notice for the circular polarization case that when the spin quantization axis is rotated from the $z$-axis to the -$\hat{\vec{B}}(\eta_r=0)$-direction, the tunneling asymmetry parameter and the spin polarization asymmetry parameter interchange their roles except a sign difference. Namely, $\mathcal{A}_{t}^{z\,\, (d)} = \mathcal{A}_{t}^{y\,\, (d)} = 0$, and $\mathcal{A}_{p}^{z\,\,(d)} = - \mathcal{A}_{t}^{y\,\, (d)} = \rho^3$.

Comparison of s-cc-SFA and d-cc-SFA for strong fields, where the spin effects are not negligible, are presented in Table~\ref{strong_field_table} in a compact way. Thus, the standard and dressed SFA give different answers for spin asymmetries and the question is which result has physical implication. In the next section we develop a simpleman model for the spin dynamics in tunnel-ionization which allows us to identify the origin of the difference between s-cc-SFA and d-cc-SFA, and to recognize the d-cc-SFA result as the physical relevant one. Note that the same conclusion is reached in \cite{Klaiber2014spin} from the comparison of analytical results with numerical simulations in the case of a linearly polarized laser field.

\section{Simpleman model for the spin dynamics} \label{sec:intuitive}

\begin{figure}[b]
  \centering
  \includegraphics[width=\linewidth]{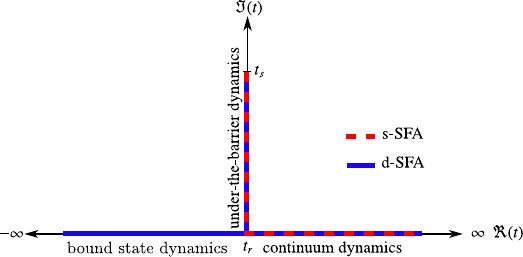}
  \caption{(Color online) The simpleman model for the time evolution of the initial bound spin state. The propagation can be split up   to three parts; the bound state, the under-the-barrier, and the continuum dynamics. The  standard SFA (red dashed curve) neglects the former, whereas the dressed SFA (blue solid curve) takes the laser field into account in   the bound state dynamics.}
  \label{fig:evo_for_simple}
\end{figure} 

All the SFA results calculated in the previous section concerning the spin dynamics can be intuitively inferred in the following simpleman model. We will describe by the simple man model spin transitions taking place at an arbitrary instant of ionization.

The complete propagation of the tunnel-ionized state can be decomposed into three parts as
\be
U(\infty, -\infty) = U_C (\infty, t_r) U_T (t_s+t_r,t_r) U_B (t_r, -\infty) \, ,
\ee
where $U_B(t_r, -\infty)$ describes the propagation of the bound state, $U_T (t_s+t_r,t_r)$ the under-the-barrier dynamics, and $U_C (\infty, t_r)$ the continuum dynamics, see Fig.~\ref{fig:evo_for_simple}. Here, $t_r$ is an arbitrary instant of ionization, and $t_s = i \sqrt{2 I_p} / E_0 $ is the imaginary Keldysh time given by Eq.~(\ref{saddle_point}), which defines how long the electron has to travel for the width of the tunneling barrier in imaginary time and determines the tunneling rate.

Firstly, the spin resolved continuum evolution can be found via the spin resolved prefactor of the Volkov wave function~(\ref{Volkov_gmg}) in the following way. The Volkov bispinors in two different times are connected to each other as
\bal
& \left[ \mathbf{I} + \frac{1}{2 c \lambda} \left( \mathbf{I} + \hat{\vec{k}} \cdot \vec{\alpha} \right)
\vec{A}(t) \cdot \vec{\alpha} \right] v_s \\ 
\nonumber & = U_C (t,t') \left[ \mathbf{I} + \frac{1}{2 c \lambda} \left( \mathbf{I} + \hat{\vec{k}} \cdot \vec{\alpha} \right)
\vec{A}(t') \cdot \vec{\alpha} \right] v_s \, ,
\eal 
with the identity matrix $\mathbf{I}$. Hence, the continuum time evolution operator yields
\be
U_C (t,t') =  \mathbf{I} + \frac{1}{2 c \lambda} \left( \mathbf{I} + \hat{\vec{k}} \cdot \vec{\alpha} \right)
\left[ \vec{A}(t) - \vec{A}(t') \right] \cdot \vec{\alpha} \, ,
\ee
which reduces to
\be
U_C (\infty, t_r) =  \mathbf{I} - \frac{1}{2 c \lambda} \left( \mathbf{I} + \hat{\vec{k}} \cdot \vec{\alpha} \right)
\vec{A}(t_r) \cdot \vec{\alpha}  \, ,
\ee
where we use the fact that $\vec{A} (\infty) \rightarrow 0$. Moreover, we will consider only the large spin components, therefore,  moving from the bispinor- to spinor-description
\be
\label{continuum_pro} U_C (\infty, t_r)   \approx  \mathbf{I} - i \frac{\xi}{2} \left[ \sin(\omega t_r) \sigma_y + \zeta \cos(\omega t_r) \sigma_x \right]  \, , 
\ee
where we further set $\lambda = 1$ for the spinor-description consistency.

\begin{figure}
  \centering
  \includegraphics[width=0.9\linewidth]{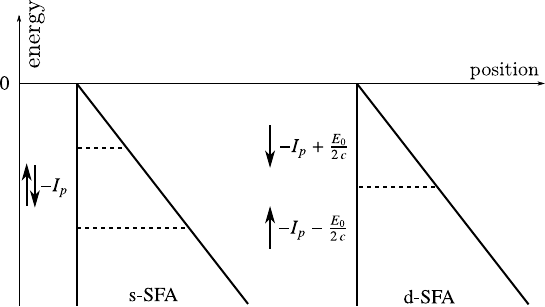}
  \caption{Intuitive description of the spin dynamics under-the-barrier can be modeled with a triangular potential barrier. In the standard SFA case, the spin states of the tunneling particle enter the barrier with the same energy $-I_p$ and are split during the under-the-barrier dynamics, whereas in the dressed SFA, first the energy of the spin state is split due to the Zeeman effect in the bound state, but later it is compensated by the energy splitting in the under-the-barrier dynamics.}
  \label{fig:box}
\end{figure}

In order to model the time evolution for the under-the-barrier dynamics, we consider a triangular barrier depicted in Fig.~\ref{fig:box} and apply WKB approximation. The WKB propagator can be mimicked by the following spin resolved propagator, which is derived via including the spin interaction Hamiltonian into Eq.~(\ref{contracted_action});
\be
U_T (t_s) = \exp\left\{i  \int_{t_r}^{t_r+t_s} dt\left[ \frac{\vec{q}(t)^2}{2} - \varepsilon_{\pm} + \frac{\vec{\sigma}\cdot \vec{B}(t)}{2 c} \right] \right\} \, ,
\label{U_T}
\ee
with $\varepsilon_{\pm}$ being the energy of the tunneling spin-$1/2$ particle, which is different in the standard and dressed SFA.

Since the bound state dynamics varies, we will investigate it in a more detail way for the standard and dressed SFA, respectively.

\subsection{Standard SFA}  \label{simpleman_SSFA}

In the standard SFA,  the influence of the laser field on the bound state is neglected. Due to the latter, all the results which we derive are valid also for an arbitrary elliptical polarization.

The evolution of the bound state between two different times in this case is given by the usual propagator as
\be
\label{trivial_bound_state_pro}
U_B (t,t') = \exp\left[- i \varepsilon_{\pm} (t-t')\right] = \exp\left[i I_p (t-t')\right] \, ,
\ee
with the energy of the tunneling bound state particle $\varepsilon_{\pm} = - I_p$, which leads to a trivial phase. Accordingly, as there is no Zeeman splitting of the energy in the bound state, the under-the-barrier propagator reads
\be
U_T (t_s) = \sqrt{W(t_r)} \exp\left[  i  \int_{t_r}^{t_r+t_s} dt  \frac{\vec{\sigma}\cdot \vec{B}(t)}{2 c} \right]  \, ,
\ee
where the tunneling probability amplitude $\sqrt{W(t_r)}$ arises from the two first terms in Eq. (\ref{U_T}), which depends on the arbitrary instant of ionization (or the electron's final momentum). For example,
\be
\sqrt{W(0)} = \exp\left(- \frac{E_a}{3 E_0}\right) \, 
\ee
for the instant of ionization that corresponds to the maximal tunneling probability. Furthermore, the spin resolved term can be written up to the order $\mathcal{O}(t_s^2)$ as
\bal
i  \int_{t_r}^{t_r+t_s} dt  \, \frac{\vec{\sigma}\cdot \vec{B}(t)}{2 c}  &\approx  i \frac{E_0 t_s}{2c} \left(  \zeta \sigma_x  \sin(\omega t_r) - \sigma_y \cos(\omega t_r) \right) \, , \\
& = \frac{\rho}{2} \left( \sigma_y \cos(\omega t_r) - \zeta \sigma_x  \sin(\omega t_r) \right) \, .
\eal

As a consequence, the final spin state in terms of the initial spin state $ \ket{\pm; \vec{\hat{s}}} \equiv \ket{\pm (-\infty); \vec{\hat{s}} } $ with an arbitrary spin quantization axis $\vec{\hat{s}} $ can be written as
\bal
\nonumber \ket{\pm (\infty) ; \vec{\hat{s}}} & = \sqrt{W(t_r)} \left[\mathbf{I} - i \frac{\xi}{2} \left( \sin(\omega t_r) \sigma_y + \zeta \cos(\omega t_r) \sigma_x \right)  \right] \\
& \times \exp\left[ \frac{\rho}{2} \left( \sigma_y \cos(\omega t_r) - \zeta \sigma_x  \sin(\omega t_r) \right) \right] \ket{\pm ; \vec{\hat{s}} } \, ,
\eal
and the transition amplitude can be given by 
\bal
\label{simpleman_s_sfa}
M_{s \rightarrow s'} &= \bra{s'}   \left[\mathbf{I} - i \frac{\xi}{2} \left( \sin(\omega t_r) \sigma_y + \zeta \cos(\omega t_r) \sigma_x \right)  \right] \\
\nonumber & \times \exp\left[ \frac{\rho}{2} \left( \sigma_y \cos(\omega t_r) - \zeta \sigma_x  \sin(\omega t_r) \right) \right] \ket{s}
\, ,
\eal
where we omit the tunneling probability amplitude for the sake of simplicity.

Let us first consider the case when the spin quantization axis is chosen along the $z$-direction. The transition amplitudes up to the leading order in $\rho$ can be found as
\begin{subequations}
\bal
M_{\pm \rightarrow \pm}^{z} & = 1 + \frac{1}{8} \xi  \rho  \left[\pm 2\zeta +i (\zeta^2 -1) \sin(2 \omega t_r) \right]\, ,\label{k-s-tun} \\ 
M_{\pm \rightarrow \mp}^{z} & = \frac{1}{2} \left[(\pm \xi -\zeta \rho ) \sin (\omega t_r)-i (\zeta  \xi \mp \rho ) \cos (\omega t_r )\right]  \, . \label{k-s-flip}
\eal
\end{subequations}
For the spin flip, we derive
\be
\label{sf_s_SFA_k_direc_simpleman}
\mathcal{F}_\pm^{z\, \, (s)} (t_r) = 1 - \frac{4(2 \pm \zeta \xi \rho)}{8 + (1+\zeta^2)\xi^2 - (1-\zeta^2)\xi^2 \cos(2 \omega t_r)} \, .
\ee
For the instant associated to the maximal tunneling probability, $t_r=0$ we have
\be
\mathcal{F}_\pm^{z\, \, (s)} (0)= \frac{\zeta \xi (\zeta \xi \mp 2 \rho)}{4+\zeta^2 \xi^2} \, ,
\ee
which coincides with the s-cc-SFA result given by Eq.~(\ref{sf_s_SFA_k_direc}). If we go further in the next orders of $\rho$, the spin flip at $t_r =0$ can be written as
\be
\label{sf_s_SFA_k_direc_simpleman_instant}
\mathcal{F}_\pm^{z\, \, (s)} (0) \approx \frac{(\zeta \xi \mp \rho)^2}{4+\zeta^2 \xi^2} \, .
\ee
Eq.~(\ref{sf_s_SFA_k_direc_simpleman_instant}) shows that both the spin precession in continuum (the term $\sim \xi$) and the spin dependent tunneling probability (the term $\sim \rho$) are contributed for the spin flip effect. In the case of linear polarization, $\zeta =0$, Eq.~(\ref{sf_s_SFA_k_direc_simpleman_instant}) gives $\mathcal{F}_\pm^{z\, \, (s)} (0) \approx \rho^2/4$, indicating spin flip only due to the tunneling. However, this is specific only for $t_r = 0$. For this reason one can introduce the mean value of the spin flip over the laser's period $T_0$, which yields
\bal
\label{sf_s_SFA_k_direc_simpleman_mean}
\langle \mathcal{F}_\pm^{z\, \, (s)} \rangle  \equiv \frac{\langle W_{\pm \rightarrow \mp}^{z\, \, (s)} \rangle}{\langle W_T^{z\, \, (s)} \rangle} &=  \frac{\xi (\xi +\zeta^2 \xi \mp 4 \zeta \rho )}{8 + (1+\zeta^2)\xi^2} \, ,
\eal
with
\be
\langle W_{s \rightarrow s'}\rangle \equiv \frac{\omega}{2 \pi} \int_{-\omega/\pi}^{\omega/\pi} dt \,  W_{s \rightarrow s'} \, .
\ee
The average spin flip in some cases can be different from the instantaneous value. For instance, in $\zeta =0$ case, Eq.~(\ref{sf_s_SFA_k_direc_simpleman_mean}) provides the strong field asymptotic, $\xi \gg 1$, $\langle \mathcal{F}_\pm^{z\, \, (s)} \rangle = 1$ in contrast to $\mathcal{F}_\pm^{z\, \, (s)} (0) \approx \rho^2/4$.

At the same order the spin asymmetries can be given by
\bal
\mathcal{A}_{t}^{z\,\, (s)} (t_r)& = 0 \, , \\
\mathcal{A}_{p}^{z\,\, (s)} (t_r)& = \frac{16 \zeta \xi \rho}{8 + (1+\zeta^2)\xi^2 - (1-\zeta^2)\xi^2 \cos(2 \omega t_r)} \, .
\eal
The tunneling asymmetry parameter is negligible for all possible intensities as well as laser's polarization similar to Eq.~(\ref{at_s_SFA_k_direc}). Further, it is independent from the instant of ionization. The polarization asymmetry parameter disappears for both of the weak field and the strong field regimes. Nonetheless, the polarization asymmetry parameter depends on the ionization moment for intermediate fields. For the maximal tunneling probability, it reads
\be
\mathcal{A}_{p}^{z\,\, (s)} (0)= \frac{2  \zeta \xi \rho}{1+ \zeta^2 \xi^2 / 4}  \, ,
\ee
which agrees with Eq.~(\ref{ap_s_SFA_k_direc}), whereas its mean value over the laser's period can be calculated as
\be
\langle \mathcal{A}_{p}^{z\,\, (s)} \rangle = \frac{2 \zeta \xi \rho}{1 + (1+ \zeta^2) \xi^2 /8} \, ,
\ee
which has the same qualitative behavior as $\mathcal{A}_{p}^{z\,\, (s)} (0)$.

In a similar way, we can deduce by the simpleman model the spin flip as well as the spin asymmetries for different spin quantization axes. For example, when we align the quantization axis along the $y$-direction, we provide the following expressions up to the order $\mathcal{O}(\rho)$
\bal
\label{sf_s_SFA_b_direc_simpleman}
\mathcal{F}_\pm^{y \, \, (s)} (t_r) &=  \frac{2\zeta^2  \xi^2 \cos (\omega t_r) (\cos (\omega t_r)\pm\rho)}{8 + (1+\zeta^2)\xi^2 - (1-\zeta^2)\xi^2 \cos(2 \omega t_r)} \, , \\
\mathcal{A}_{t}^{y\,\, (s)}  (t_r) & = 2 \rho \cos(\omega t_r) \, , \\
\mathcal{A}_{p}^{y\,\, (s)} (t_r) & = 2 \rho \cos(\omega t_r)  \\
\nonumber & \times \left[1- \frac{4 \zeta^2 \xi^2}{8 + (1+\zeta^2)\xi^2 - (1-\zeta^2)\xi^2 \cos(2 \omega t_r)} \right] \, .
\eal
For the linear polarization case, the spin flip vanishes and the both asymmetries simplify to $\mathcal{A}_{t}^{y\,\, (s)} (t_r)= \mathcal{A}_{t}^{y\,\, (s)} (t_r)= 2 \rho \cos(\omega t_r)$. The spin asymmetries  as well as the spin flip for a nonvanishing polarization depend on the instant of ionization even for strong fields. While we derive
\bal
\mathcal{F}_\pm^{y \, \, (s)} (0) & = \frac{(1\pm \rho) \zeta^2 \xi^2}{4 + \zeta^2 \xi^2}   \, , \\
\mathcal{A}_{t}^{y\,\, (s)} (0)& = 2\rho \, , \\
\mathcal{A}_{p}^{y\,\, (s)} (0)& = \frac{2 \rho  \left(4-\zeta ^2 \xi ^2\right)}{4+\zeta ^2 \xi ^2} \, , 
\eal
for the maximal tunneling probability, and we notice that the simpleman results capture the corresponding SFA results, Eqs.~(\ref{sf_s_SFA_b_direc})-(\ref{ap_s_SFA_b_direc}), their mean value can be calculated as
\bal
\langle \mathcal{F}_\pm^{y \, \, (s)} \rangle  & = \frac{\zeta ^2 \xi ^2}{8 + (1+\zeta^2)\xi^2} \, , \\
\langle \mathcal{A}_{t}^{y\,\, (s)} \rangle & = 0 \,, \\
\langle \mathcal{A}_{p}^{y\,\, (s)} \rangle & = 0 \, .
\eal
One should note that the time averaging destroy the asymmetries. The spin flip qualitatively remains the same at the time averaging, but decreases two times.

If the quantization axis is along the $x$-direction, at the leading order of $\rho$, the spin flip and the spin asymmetries can be written as
\bal
\label{sf_s_SFA_e_direc_simpleman}
\mathcal{F}_\pm^{x \, \, (s)}  (t_r)&  =   \frac{\xi^2 \sin(\omega t_r) (\sin(\omega t_r) \mp \zeta \rho)}{4 + \xi^2 \left(\zeta^2 \cos^2(\omega t_r) + \cos^2(\omega t_r)\right)} \, , \\
\mathcal{A}_{t}^{x\,\, (s)}  (t_r)& = -2 \zeta \rho  \sin (\omega t_r) \, , \\
\mathcal{A}_{p}^{x\,\, (s)} (t_r)&  = -2 \zeta \rho  \sin (\omega t_r) \\
\nonumber & \times \left[1- \frac{4 \xi^2}{8 + (1+\zeta^2)\xi^2 - (1-\zeta^2)\xi^2 \cos(2 \omega t_r)} \right] \, .
\eal
All the spin effects depend on the instant of ionization, and all of them disappear for the maximum tunneling probability at the leading order in $\rho$. If we further go the next order, the spin flip for the maximal tunneling probability can be given by
\be
\mathcal{F}_\pm^{x \, \, (s)} (0) = \frac{\rho^2}{4}  \, ,
\ee
which is consistent with Eq.~(\ref{sf_s_SFA_e_direc}).
The mean values over the laser's period, on the other hand, can be given by
\bal
\langle \mathcal{F}_\pm^{x \, \, (s)} \rangle  & = \frac{\xi^2}{8 + (1+\zeta^2)\xi^2} \, , \\
\langle \mathcal{A}_{t}^{x\,\, (s)} \rangle & = 0 \, , \\
\langle \mathcal{A}_{p}^{x\,\, (s)}  \rangle & = 0 \, .
\eal
In this case, the average value of the spin flip is quite different than the instantaneous value. Thus $\langle \mathcal{F}_\pm^{x \, \, (s)} \rangle  \approx 1$ at $\xi \gg 1$.

We notice that our simpleman model captures the main features of the spin dynamics and is able to reproduce the results of the SFA. The averaging of the spin flip and asymmetries over the ionization time in some cases modifies significantly the result.

\subsection{Dressed SFA} \label{simpleman_DSFA}

In the dressed SFA, the scenario is  changing because we take into account the  bound state evolution in the laser field. The bound state propagation is not a trivial phase in this case, but includes the spin  precession in the bound state. Accordingly, the Hamiltonian for the bound state propagation can be represented as
\be
\label{simpleman_hamiltonian}
H = H_0 \mathbf{I} + \frac{\vec{\sigma}\cdot \vec{B}(t)}{2 c} \, ,
\ee
where $H_0$ is the usual atomic Hamiltonian of the electron in the Coulomb field of the core. We would like underline that once the solution of the Schr\"{o}dinger equation for the Hamiltonian~(\ref{simpleman_hamiltonian}) is known, the developed simpleman model is valid for any polarization, and accordingly we discuss the linear and circular polarization cases separately for the dressed SFA.

In addition to the trivial phase~(\ref{trivial_bound_state_pro}), in this scenario the bound state propagator for a linearly polarized field can be calculated as
\bal
\label{UB_linear}
U_B (t_r, - \infty) & = \exp\left( - i \int_{-\infty}^{t_r} \frac{\sigma_y B_y (t)}{2 c} \right) \, , \\
& = \cos\left(\frac{\xi}{2} \sin(\omega t_r)\right)\mathbf{I} + i \sigma_y \sin\left(\frac{\xi}{2} \sin(\omega t_r)\right) \, .
\eal

Furthermore, due to the Zeeman splitting, the energy of the tunneling bound spin states whose quantization axis is along $y$-direction will read $\varepsilon_{\pm} =  - I_p \mp  E_0 \cos(\omega t_r)/ (2c)$ for a certain instant of ionization $t_r$, where $\varepsilon_{+ (-)}$ is for spin-up (down) state. Consequently, as the spin states having different quantization axes can always be expanded by the basis vectors $\ket{\pm; \hat{\vec{y}}}$, the Zeeman splitting and the spin interaction under-the-barrier cancel each other, and the propagator is simply given by the tunneling probability amplitude
\be
\label{d-U_T}
U_T (t_s) = \sqrt{W(t_r)} \, .
\ee
We conclude that the tunneling probability is spin independent in the dressed SFA  at the leading order in $\rho$. Therefore, the spin asymmetries disappear at this order [the contribution of higher orders are discussed in the next subsection].

After including the continuum propagation for $\zeta = 0$, the transition amplitude in the case of the dressed SFA for linearly polarized field becomes
\bal
& M_{s \rightarrow s'}  = \times \bra{s'}\left( \mathbf{I} - i \frac{\xi}{2 \lambda} \sin(\omega t_r) \sigma_y  \right)  \\
\nonumber & \times \left( \cos\left(\frac{\xi}{2} \sin(\omega t_r)\right)\mathbf{I} + i \sigma_y \sin\left(\frac{\xi}{2} \sin(\omega t_r)\right) \right) \ket{s} \, .
\eal

First of all, the transition amplitude is a function of $\sigma_y$, as a consequence there cannot exist a spin flip when the spin quantization axis is the $y$-direction, which explains the derived result~(\ref{sf_d_SFA_b_direc}). Furthermore, due to the symmetry reason, the spin flip for the case when the spin quantization axis is the $z$-direction is the same as when it is the $x$-direction, and they can be calculated as
\bal
\label{sf_d_SFA_e_direc_simpleman}
& \mathcal{F}_\pm^{z,x \, \, (d)}  (t_r) =  \frac{ 2 \left[ \xi  \sin(\omega t_r)  \cos\left(\frac{\xi \sin(\omega t_r)}{2}  \right)  -2  \sin\left(\frac{\xi \sin(\omega t_r)}{2}  \right) \right]^2}{8 + \xi^2 (1-\cos(2\omega t_r))} \, .
\eal
At $t_r = 0$, $\mathcal{F}_\pm^{z,x \, \, (d)}  (0) = 0$. If $t_r \ne 0$, $\mathcal{F}_\pm^{z,x \, \, (d)}  (t_r) \sim \xi^2$ for weak fields, and it is negligible. However, its strong field limit can be given by
\be
\mathcal{F}_\pm^{z,x\, \,  (d)} = \cos^2\left(\frac{\xi}{2} \sin(\omega t_r) \right) \, .
\ee

As we discussed in Sec.~\ref{sec:bound_state}, the bound spin highly oscillates between up and down states in a very short time interval for strong fields. As a result, it is physically more correct to present the mean value of the spin flip instead of its instantaneous value at $t_r=0$. Furthermore, one can obtain the mean value corresponding to the maximal tunneling probability by averaging over a period $T$ fulfilling the condition~(\ref{period_condition}), and it is
\be
\label{linear_strong_dsfa}
\langle \mathcal{F}_\pm^{z,x \, \,  (d)} \rangle_T = \frac{1}{2} \, ,
\ee
which agrees with Eq.~(\ref{sf_d_SFA_k_direc}), and Eq.~(\ref{sf_d_SFA_e_direc}) as well as with Ref.~\cite{Klaiber2014spin}. Note that the mean value over the laser period $T_0$ needs a numerical calculation.

For the circular polarization case, we first notice that solving the coupled differential equations~(\ref{ce}) corresponds to solving the Schr\"{o}dinger equation for the Hamiltonian $H = \vec{\sigma}\cdot \vec{B}(t)/(2 c/\delta) $. Furthermore, the corresponding propagator can be written as
\be
U_B (t_r, - \infty) = \begin{pmatrix}
            C^{++} (t_r)  &  C^{-+} (t_r) \\ 
	    C^{-+} (t_r)  &  C^{--} (t_r) \\ 
          \end{pmatrix} \, ,
\ee
with the coefficients~(\ref{circ_at_instant}). As a consequence, the bound state propagator
for a circularly polarized field can be obtained with the replacement of $\delta \rightarrow 1$ in the coefficients~(\ref{circ_at_instant}), which reads
\bal
\label{UB_circ}
& U_B (t_r, - \infty) = \frac{1}{\sqrt{\xi ^2+(1+ \sqrt{1+\xi^2})^2}}  \\
\nonumber & \times \begin{pmatrix}
            \left( 1+\sqrt{1+\xi^2} \right)e^{-i \omega t_r \frac{1-\sqrt{1 + \xi^2}}{2}}  &  i \xi e^{-i \omega t_r \frac{1 + \sqrt{1 + \xi^2}}{2}} \\ 
	    i \xi e^{i \omega t_r \frac{1 + \sqrt{1 + \xi^2}}{2}}  & \left( 1+\sqrt{1+\xi^2} \right) e^{i \omega t_r \frac{1-\sqrt{1 + \xi^2}}{2}} \\ 
          \end{pmatrix} \, .
\eal
While the under-the-barrier propagator is the same as Eq.~(\ref{d-U_T}), the continuum propagator for the circular polarization case reads
\be
U_C (\infty, t_r) = 1 - i \frac{\xi}{2} \left[ \sin(\omega t_r) \sigma_y +  \cos(\omega t_r) \sigma_x \right] \, .
\ee
Thereby, we can give an intuitive description of the spin effects for a circularly polarized field. For instance, when the spin quantization axis is along the $z$-direction, the spin flip is
\bal
& \mathcal{F}_\pm^{z \, \, (d)}  (t_r) =  \frac{\xi^2 \left[4 - 4 (1+ \Xi)Cos\left(\omega t_r (1- \Xi)\right) + (1+\Xi)^2\right]}{2(4+\xi^2)\left(\Xi^2 +\Xi \right)} \, .
\eal
with $\Xi \equiv \sqrt{1+\xi^2}$. For a weak field the spin flip is negligible, whereas in the strong field, $\xi \gg 1$ we derive
\be
\mathcal{F}_\pm^{z \, \, (d)}  (0) = \frac{1}{2} \, ,
\ee
for the instantaneous value at $t_r=0$. The mean value of the spin flip over the period $T$~(\ref{period_condition}),
\be
\langle \mathcal{F}_\pm^{z \, \,  (d)} \rangle_T = \frac{1}{2} \, ,
\ee
which coincides with its instantaneous value. This result is a consequence of this particular choice of quantization axis as we discussed in Sec.~\ref{bound_state_dynamics_circ}. In a similar way, the strong field limit of the spin flip for the other spin quantization axes can be derived as
\bal
\mathcal{F}_\pm^{y \, \, (d)}  (t_r) & = \frac{1}{2} \left[1 - \frac{\cos\left[ \omega t_r (1-\Xi) \right]-\cos\left[ \omega t_r (3-\Xi) \right]}{2} \right] \, , \\
\mathcal{F}_\pm^{x \, \, (d)}  (t_r) & = \frac{1}{2} \left[1 - \frac{\cos\left[ \omega t_r (1-\Xi) \right]+\cos\left[ \omega t_r (3-\Xi) \right]}{2} \right] \, .
\eal
While at $t_r = 0$, we provide
\bal
\mathcal{F}_\pm^{y \, \, (d)}  (0) & = \frac{1}{2} \, , \\
\mathcal{F}_\pm^{x \, \, (d)}  (0) & = 0  \, ,
\eal
their mean values over the period $T$ are
\bal
\langle \mathcal{F}_\pm^{y \, \,  (d)} \rangle_T & = \frac{1}{2} \, , \\
\langle \mathcal{F}_\pm^{x \, \,  (d)} \rangle_T & = \frac{1}{2} \, ,
\eal
which are consistent with Eq.~(\ref{sf_d_SFA_b_direc}), and Eq.~(\ref{sf_d_SFA_e_direc}).

Thus, the simpleman model clearly shows that the spin dynamics in the bound state is very important for developing of spin effects during tunnel-ionization. This dynamics is included only in the dressed SFA and, therefore, the physically correct results are those predicted by the dressed SFA.

\subsection{Origin of the spin asymmetries in the dressed SFA}

If we compare the results developed in the previous sections~\ref{simpleman_SSFA} and \ref{simpleman_DSFA}, we notice that the origin of the spin asymmetries in the standard SFA is the spin interaction under-the-barrier, which leads to different tunneling probabilities for spin-up and spin-down states. Since the Zeeman splitting in the bound state compensates this spin interaction in the dressed SFA, there is no spin asymmetries at the leading order of $\rho$. However, the magnetic field in the rest frame of the electron is slightly different in the bound state and under-the-barrier, which have an impact on the Zeeman splitting compensation and can lead to a spin asymmetry. This can be described by improving our simpleman model in the following way.

If we go further in the Foldy-Wouthuysen expansion of the Hamiltonian, we can write down the propagator for the under-the-barrier dynamics as
\bal
\label{fw_propagator_for_tunneling}
& U_T (t_s)  \\
\nonumber & = \resizebox{.9\hsize}{!}{$\exp\left\{i  \int_{t_r}^{t_r + t_s} dt\left[ \frac{\vec{q}(t)^2}{2} - \varepsilon_{\pm} + \frac{\vec{\sigma}\cdot \vec{B}(t)}{2 c}
 + \frac{\vec{\sigma}\cdot(\vec{E}(t) \times \vec{q}(t)) }{4 c^2} \right] \right\}$} \, ,
\eal
where the last term in the exponent of Eq.~(\ref{fw_propagator_for_tunneling}) describes the effect of the magnetic field in the rest frame of the electron during tunneling. Note that as the momentum vanishes for the bound state, i.e., $\langle \vec{p} \rangle_B = 0$, we do not need to modify the energy of the tunneling bound spin states. Moreover, the rest frame effect depends on the final momentum distribution via the term $\vec{E}(t) \times \vec{q}(t)$. Thereby, in order to elaborate the spin asymmetries in the dressed SFA, we consider the instant of ionization associated to the maximal tunneling probability.  The rest frame effect term, then, can be calculated up to the order $\mathcal{O}(t_s^2)$ as
\bal
\nonumber i \int_0^{t_s} dt \, \frac{\vec{\sigma}\cdot(\vec{E}(t) \times \vec{q}(t)) }{4 c^2} & \sim i \frac{\sigma_y E_0 q_z (0)  t_s}{4c^2} \, , \\
& = \frac{\sigma_y \rho^3}{12} \, ,
\eal
where we have further used $q_z (t) = - 2I_p /(3c)$ via Eq.~(\ref{mom_q}). As a result, the under-the-barrier propagator yields
\be
\label{under_the_barrier_prop_improved}
U_T (t_s)  = \exp\left( - \frac{E_a}{3 E_0} \right) \exp\left( \frac{\sigma_y}{12} \rho^3\right) \,
\ee
for the maximal tunneling probability.

Combining the bound state propagator for linear~(\ref{UB_linear}) and circular polarizations~(\ref{UB_circ}) together, we derive
\bal
& U_B (0, - \infty) = (1-\zeta)\mathbf{I} +
\frac{\zeta \left[ \left( 1+\sqrt{1+\xi^2} \right)\mathbf{I} + i \xi \sigma_x \right] }{\sqrt{\xi ^2+(1+ \sqrt{1+\xi^2})^2}}  \, ,
\eal
with $\zeta = \{0, 1\}$. Including the continuum propagator
\be
U_C (\infty, 0) =  \mathbf{I} - i \frac{\xi \zeta}{2} \sigma_x \, , 
\ee
the transition amplitude can be written as
\bal
M_{s \rightarrow s'} &= \bra{s'}  \left[ \left( \mathbf{I} - i \frac{\xi \zeta}{2} \sigma_x \right) \left( \mathbf{I} + \frac{\sigma_y}{12}\rho^3  \right) \right.  \\ 
\nonumber & \times \left. \left( (1-\zeta)\mathbf{I} +
\frac{\zeta \left[ \left( 1+\sqrt{1+\xi^2} \right)\mathbf{I} + i \xi \sigma_x \right] }{\sqrt{\xi ^2+(1+ \sqrt{1+\xi^2})^2}} \right) \right]  
\ket{s}  \, ,
\eal
where an expansion over a small parameter $\rho$ is used. Then, in the linear polarization case, $\zeta = 0$ the transition amplitude yields
\be
M_{s \rightarrow s'} = \bra{s'}  \left(\mathbf{I} + \frac{\sigma_y}{12}\rho^3  \right) \ket{s} \, .
\ee
Since the amplitude depends only on $\sigma_y$, there cannot exist the spin asymmetries when the spin  quantization axis is along the $z$ or $x$-directions. This is the underlying reason of Eqs.~(\ref{at_d_SFA_k_direc}) and (\ref{ap_d_SFA_k_direc}) as well as Eq.~(\ref{atp_d_SFA_e_direc}). However, when the spin quantization direction is chosen in the $y$-direction, we derive
\be
\mathcal{A}_{t}^{y \,\, (d)} = \mathcal{A}_{p}^{y \,\, (d)} = \frac{\rho^3}{3} \, ,
\ee
which agrees with Eqs.~(\ref{at_d_SFA_b_direc}) and (\ref{ap_d_SFA_b_direc}) up to $1/3$ prefactor.

In the case of a circularly polarized field, the corresponding spin asymmetries can be written as 
\begin{subequations}
\bal
\mathcal{A}_{t}^{z \,\, (d)} & = \frac{\xi}{3\sqrt{1+\xi^2}}\rho^3 \, , \quad \mathcal{A}_{p}^{z \,\, (d)} = \frac{4\xi}{3(4+\xi^2)}\rho^3 \, , \\
\mathcal{A}_{t}^{y \,\, (d)} & = \frac{1}{3\sqrt{1+\xi^2}}\rho^3 \, , \quad \mathcal{A}_{p}^{y \,\, (d)}  = \frac{4-\xi^2}{3(4+\xi^2)}\rho^3 \, , \\
\mathcal{A}_{t}^{x \,\, (d)} & = 0 \, , \hspace{1.9cm} \mathcal{A}_{p}^{x \,\, (d)} = 0 \, ,
\eal
\end{subequations}
and their strong field limit, $\xi \gg 1$ is
\begin{subequations}
\bal
\mathcal{A}_{t}^{z \,\, (d)} & = \frac{\rho^3}{3} \, , \quad \mathcal{A}_{p}^{z \,\, (d)} = 0 \, , \\
\mathcal{A}_{t}^{y \,\, (d)} & = 0 \, , \quad \mathcal{A}_{p}^{y \,\, (d)}  = -\frac{\rho^3}{3} \, , \\
\mathcal{A}_{t}^{x \,\, (d)} & = 0 \, , \quad \mathcal{A}_{p}^{x \,\, (d)} = 0 \, .
\eal
\end{subequations}
Similar to the linear polarization case they agree with the d-cc-SFA results up to the $1/3$ factor (see Table~\ref{strong_field_table}). 

As a conclusion, the simpleman model describes correctly the qualitative behavior of the asymmetry parameters. Here we would like stress that we divide the total propagator into three parts: the bound, under-the-barrier, and continuum parts. While the bound and continuum propagators are unitary operators, the tunneling asymmetry parameter $\mathcal{A}_{t}$ vanishes for these propagations. The latter follows from the fact that since the conservation of the probability implies $W_{+ \rightarrow +} + W_{+ \rightarrow -}= 1$ as well as $W_{- \rightarrow +} + W_{- \rightarrow -} = 1$, with $W_{s \rightarrow s'} = \left| \bra{s'} U(t, t' ) \ket{s} \right|^2$, by definition $\mathcal{A}_{t}$ essentially disappears. Nonetheless, the under-the-barrier evolution operator is not a unitary operator as it determines the tunneling rate. Therefore, the asymmetries are originated from the tunneling step.

\subsection{Further contributions to spin effects}

Finally, we would like to discuss further contributions to the spin effects that have been neglected in our treatment. First of all, we would like to emphasize that the following relativistic corrections, $\lambda \rightarrow 1 -\rho^2 / 6$, $\delta \rightarrow 1 - \rho^2 / 3 $, and $t_s \rightarrow i \sqrt{2 I_p}/E_0 ( 1 -5 \rho^2 / 72 )$ as well as the effect of the bi-spinor description can improve the accuracy of the results of the simpleman model.

Next, we have neglected the spin-orbit interaction during the electron motion in the continuum under the action of the laser and Coulomb fields.  In the  paper II, we have identified the spin-orbit coupling term in the continuum wave function. Including this term  in the continuum propagator will result in a new term for the matrix element describing spin transitions
\be
\frac{1}{2c \lambda} \int_t^\infty d t' \vec{\alpha} \cdot \vec{\nabla} V^{(C)} (r(t')) \, .
\ee
This will modify not only the simpleman model but also the SFA calculations.

Finally, there could be a relation of spin effects to the tunneling delay time \cite{Yakaboylu_2014_td}, which needs investigation beyond the quasiclassical description of tunneling. For instance, in the simpleman model the time evolution of the wave function will be modified as
\be
U(\infty, -\infty) = U_C (\infty, t_r + \tau) U_T (t_r + t_s + \tau, t_r) U_B (t_r, -\infty) \, ,
\ee
if we introduce a real and positive time $\tau$ associated to the time spent under-the-barrier, which would lead to an additional precession and spin effect modifications. 

The mentioned modifications for the spin transition probabilities  will be discussed elsewhere.

\section{Experimental observability}\label{sec:experiment}

\subsection{Detection of photoelectrons}

\begin{figure}
  \centering
  \includegraphics[width=0.8\linewidth]{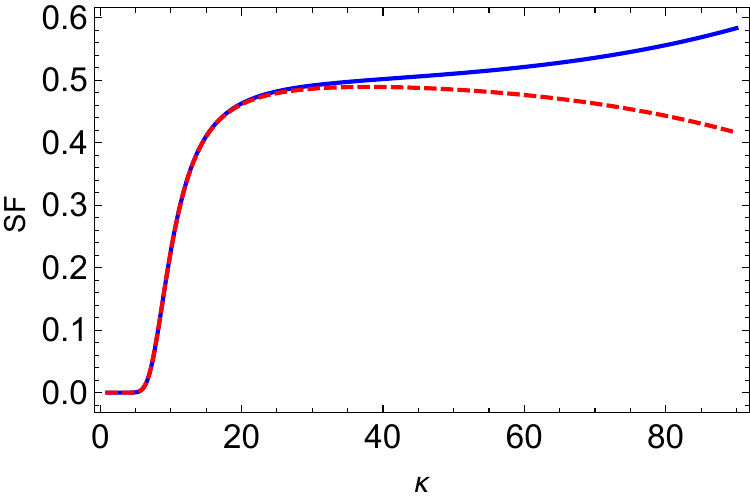}
  \caption{(Color online) The spin flip associated to the maximal tunneling probability for a circularly polarized field when the spin quantization axis is the laser's propagation direction within d-cc-SFA. The blue solid and the red dashed lines represent $\mathcal{SF}_{+}^{z\,\, (d)}$ and $\mathcal{SF}_{-}^{z\,\,(d)}$, respectively. The applied parameters are $\omega=0.05$, and $E_0/E_a = 1/30$.}
  \label{fig:SF_single}
\end{figure}

The measurement of the photoelectron spin flip during tunnel-ionization requires an initially polarized atomic target  and a detection of the photoelectron spin polarization. 
The photoelectron spin polarization can be measured using Mott polarimetry \cite{Kessler_1985,Tioukine_2011,Jakubassa-Amundsen_2012}. The latter is based on the left-right asymmetry of Mott scattering cross-section on a high-$\kappa$ target. It depends on the electron spin polarization ${\cal P}_{\bot}$, transverse to the scattering plane; $d\sigma=d\sigma_0[1+ {\cal P}_{\bot} S(\theta)]$, where $ d\sigma$ and $d\sigma_0$ are the spin resolved and spin-averaged cross-sections of Mott scattering, respectively, $S(\theta)$ is the so-called Sherman function, the maximum of which is approximately 0.5 in the case of a gold target at electron energies of the order of megaelectronvolts \cite{Tioukine_2011}, and $\sim 10^{-2}$ in the case of zinc or lead targets at electron energies of order of hundred megaelectronvolts \cite{Tioukine_2011}. For instance, when the spin of the ionized electron is polarized in the laser propagation direction in a circularly polarized laser field, the electron spin polarization in the final state is $1/2$ for $\xi\gg 1$, according to Table~\ref{strong_field_table} and see Fig.~\ref{fig:SF_single}. This means that the spin in the final state will be oriented transversely with respect to the propagation direction. Therefore,   ${\cal P}_{\bot} \sim 1$ is possible, because in the relativistic regime the electron final momentum is mostly along the laser propagation direction. At $\xi\sim 10$ (the laser intensity of the order of $10^{20}$ W/cm$^{2}$), ${\cal P}_{\bot}  S(\theta)_{\rm max}\approx 10^{-2}$, and the relative error of the signal ($\sim 1/\sqrt{N_s^{(t)}}$, with the total scattering events $N_s^{(t)}$) should be smaller than $10^{-2}$ to distinguish the electron polarization, i.e., $N_s^{(t)}=v_{shot} N_s^{(1)}>10^4$, where $v_{shot}$ is the number of laser shots, and $N_s^{(1)}$ is the number of scattering events per the laser shot. The latter is determined by the ionization ($W_i$) and the Mott scattering ($W_M$) probabilities $N_s^{(1)}\sim W_i\times W_M\times (\delta\theta /\pi)$, where the effective  interval of the scattering angle (where $S(\theta)\sim 10^{-2}$ \cite{Jakubassa-Amundsen_2012}) is $\delta\theta\sim 5^{\rm o}$. Estimating the Mott scattering probability via $W_M\approx (\kappa^2\pi a_B^2/\gamma_l^2) \rho_0 \ell\approx 1$, with the Bohr radius $a_B$, the target length $\ell=0.1$ ${\mu}$m,  the Lorentz factor $\gamma_l \approx \xi^2= 100$, $\kappa=30$, and the density $\rho_0=2\cdot 10^{22}$ cm$^{-3}$, each electron will produce a scattering event but the hindering multiple scattering will be avoided. 
The ionization probability per laser shot can be estimated as $W_i\sim 10^{-2}\kappa^2$ (at a fixed $E/E_a=1/10$, the laser pulse duration of 100 fs) \cite{Klaiber_2013b}. Then the scattering events per laser shot is $N_{s}^{(1)}\approx W_M\times (\delta\theta /\pi)\times W_i\sim 3\times 10^{-2}$. The number of laser shots required for the necessary signal resolution is $\nu_{shot}> N_{s}^{(1)}/ N_s\approx 3\times 10^5$, which can be realized even with 1 Hz laser system.

For typical experimental parameters, e.g., ionization of hydrogenlike Ne$^{9+}$-ions in a strong infrared laser field with an intensity of $ 10^{20}$ W/cm$^{2}$, the d-cc-SFA predicts a spin-flip relative probability of about 0.1. 

\subsection{Detection of ions}

The electron spin flip can also be revealed via a measurement of ion parameters. The angular momentum change of the ion during ionization $\Delta J_i$ can be related to the electron spin change $\Delta S$, using the angular momentum and the energy conservation laws in the case of circular polarization 
\bal
n+J_i+J_e&= J'_i+ J'_e,\\
n\omega &=I_p+2U_p,
\eal
where $n$ is the number of the absorbed photons at ionization,  $U_p=c^2\xi^2/2$ is the ponderomotive potential, $J_{i,e}$ and $J'_{i,e}$ are the ion and the electron total angular momentum before and after the ionization, respectively, $J_{i,e}= L_{i,e}+S_{i,e}$ with $L_{i,e}$ and $S_{i,e}$ being the orbital angular momentum and the spin of the ion and electron, respectively, and $\Delta J_{i,e}=J'_{i,e}-J_{i,e}$, $\Delta S_{i,e}=S'_{i,e}-S_{i,e}$ and $n=\Delta J_e+\Delta J_i$. Taking into account that the change of the electron angular momentum during above-threshold ionization, which is
\bal
\Delta L_e\approx \frac{I_p+2U_p}{\omega}\,, 
\eal
see \cite{Reiss_1990}, the change of the electron spin can be measured via the ion angular momentum 
\bal
\Delta S=-\Delta J_i \,.
\eal
Let us employ a circularly polarized laser field and polarize the initial electron spin along the laser propagation direction (the ion nuclear spin is assumed to be vanishing $S_i=0$). Even when the electron spin is not changed during the ionization $\Delta S_e=0$, the ion  carries out a small part of the  angular momentum provided by $n$ absorbed photons. In particular, with a right circular polarization, the ion gets an orbital angular momentum 
\bal
L_z^R=L_0\equiv \frac{m}{M}\frac{I_p+2U_p}{\omega}\,, 
\eal
where $m$ and $M$ are the electron and ion mass, respectively. The latter follows from the fact that the ion distance with respect to the center-of-mass of the ion-electron system $m/M$ times smaller than the electron distance. With a left circular polarization, the ion acquires an angular momentum $L_z^L=-L_0$. When the spin flip happens, $L_z^R=L_0-1$ and $L_z^L=-L_0-1$, respectively. Therefore, the spin flip will be indicated by a non-vanishing signal for the difference in the ion angular distribution with the left and right polarized laser field. The mentioned signal for the difference in the ion angular distribution can be estimated 
as 
\bal
{\cal S}&\sim ||Y_{L_0+1,L_0+1}|^2-|Y_{L_0-1,L_0-1}|^2| \\
&\sim |\sin^{2(L_0+1)}\theta-\sin^{2(L_0-1)}\theta|\sim 2\delta\,, \nonumber
\eal
where $Y_{l,m}$ are the spherical harmonics, $\theta$ is the ion scattering angle with respect to the laser propagation direction, and $\delta\equiv \cot^2 \theta\ll 1$. Then, the required number of laser hots is $\nu_{shot}\sim 10^6$ at $\delta\sim 10^{-3}$ (the ions are observed in the transverse to the laser propagation direction within an angle of 20 mrad).

\section{Conclusion} \label{sec:conclusion}

In this paper we have investigated the spin effects during strong field ionization in the tunneling regime for linearly as well as circularly polarized laser fields. The spin effects are fully described by two types of spin asymmetries, tunneling asymmetry and polarization asymmetry, as well as by the spin flip probability. The tunneling asymmetry describes the asymmetry in ionization of the initially polarized target, while the polarization asymmetry describes the degree of the photoelectron polarization from an unpolarized target. The spin resolved differential ionization rates as well as the spin asymmetries are derived for the maximal tunneling probability. For this purpose, the final momentum distribution of the tunnel-ionized electron is calculated. The results are further generalized to an arbitrary spin quantization axis.

Two versions of the Coulomb corrected SFA have been considered. While in s-cc-SFA the influence of the laser field on the bound state is completely neglected, the latter is properly treated in d-cc-SFA. Therefore, we have discussed the spin dynamics in the bound state  driven by linearly as well as circularly polarized fields. The physically relevant predictions for the spin effects is given by the dressed SFA. Generally, the spin effects calculated with s-cc-SFA are overestimated.  

A simpleman model for spin effects is developed which transparently shows how the spin effects arise in three steps; spin precession in the bound state, spin rotation during tunneling and further spin precession in the continuum.  

The parameter $\xi$ determines the field regime, and in the weak field regime $\xi\ll 1$ the spin flip as well as the spin asymmetries are negligible. The spin effects are considerable for strong fields $\xi \gg 1$, where the spin flip probability is about 1/2. The spin flip probability is mainly determined by the bound-state dynamics, however
the latter has no effect on the spin asymmetries. Indeed, the spin asymmetries, which increases with increasing the charge of atomic core, are a consequence of the effect of the magnetic field in the rest frame of the electron during tunneling as indicated by the developed simpleman model.

The most favorable for experimental observation of spin effects is the spin flip by using  moderate highly charged ions with a charge of the order of $\kappa\sim 20$ and a laser field with an intensity of $I\sim 10^{22}$ W/cm$^2$.

\section*{Acknowledgments}

We are grateful to C. H. Keitel, A. Di Piazza, and S. Meuren for valuable discussions, and M. Twardy for useful comments. EY would like to thank O. Skoromnik, and L. Zhang for fruitful discussions and for advices related to the Mathematica program.

\section{Appendix} \label{sec:appendix}

\subsection{Classical equations of motion of a charged particle in a plane wave} \label{ceom}

The most general gauge potential for a plane wave can be given by
\be
A^\mu = \epsilon_1^\mu f_1 (\eta) + \epsilon_2^\mu f_2 (\eta) \, ,
\ee
with the phase $\eta = k x/\omega$ and the wave vector $k^\mu$ ($k^2 = 0$). Here 
$\epsilon_1^\mu$,
$\epsilon_2^\mu$ are the polarization vectors such that $\epsilon_1^2 = 
\epsilon_2^2 = -1$ and
$\epsilon_1 k = \epsilon_2 k = \epsilon_1 \epsilon_2 = 0$. Then the field 
strength tensor $F_{\mu
\nu}$ reads
\be
F^{\mu \nu} = \epsilon_1^{\mu \nu} \dot{f_1} (\eta) + \epsilon_2^{\mu \nu} 
\dot{f_2} (\eta) \, ,
\ee
where $\epsilon_1^{\mu \nu} = k^\mu \epsilon_1^\nu - k^\nu \epsilon_1^\mu$, 
$\epsilon_2^{\mu \nu} =
k^\mu \epsilon_2^\nu - k^\nu \epsilon_2^\mu$, and dot denotes the derivative 
with respect to the
phase $\eta $.

The classical equations of motion are obtained by solving the Lorentz force law, which can be written in the proper time parametrization as
\be
\frac{d p^\mu}{d \tau} = - \frac{1}{c} F^{\mu \nu} p_\nu \, .
\ee
The relation $k_\mu \epsilon_1^{\mu \nu} = k_\mu \epsilon_2^{\mu \nu} = 0$ implies that $kp$ is a constant of motion. Using further the following identity
\be
\frac{d p^\mu}{d \tau} = \frac{d p^\mu}{d \eta} \frac{d \eta}{d \tau} = 
\dot{p}^\mu \lambda
\, ,
\ee 
with $\lambda = k p/\omega$, the Lorentz force law in the phase parametrization 
yields
\be
\label{lfl_phase}
\dot{p}^\mu = \frac{1}{c} \dot{A}^\mu - \frac{1}{\lambda \omega c} k^\mu \dot{A} 
p \, .
\ee
Moreover, the last term in the above equation can also be given by
\be
\dot{A} p = \frac{d}{d \eta} (A p) - A \dot{p} = \frac{d}{d \eta} \left( A p 
-\frac{1}{2 c} A^2
\right)\, ,
\ee
where we used the contraction of Eq.~(\ref{lfl_phase}) with $A_\mu$ after the second equality sign. Then, Eq.~(\ref{lfl_phase}) can be written as
\be
\dot{p}^\mu = \frac{d}{d \eta}  \left( \frac{1}{c} A^\mu  - \frac{1}{\lambda 
\omega c} k^\mu \left(
A p -\frac{1}{2 c} A^2 \right) \right) \, ,
\ee
whose solution in terms of the initial phase $\eta_i$ can be found as
\bal
\nonumber p^\mu (\eta) &= p^\mu (\eta_i)  +  \frac{A^\mu (\eta) -A^\mu 
(\eta_i)}{c}   \\
&- \frac{k^\mu}{\lambda \omega c}  \left( p^\nu (\eta_i)  +  \frac{A^\nu (\eta) 
-A^\nu (\eta_i)}{2
c}\right) (A_\nu (\eta) -A_\nu (\eta_i))  \, ,
\eal
where we used the following relation
\be
A_\mu (\eta) p^\mu (\eta) = A_\mu (\eta) p^\mu (\eta_i) + \frac{1}{c}A_\mu (\eta) \left( A^\mu(\eta) - A^\mu(\eta_i) \right) \, .
\ee
In the velocity gauge $A^\mu = (0, c \vec{A})$, the three-momentum and the energy, then, yield
\bal
\nonumber \vec{p}(\eta) &= \vec{p} (\eta_i)  +  \vec{A}(\eta) - \vec{A} (\eta_i) 
\\
\label{eom} &+ \frac{\hat{\vec{k}}}{\lambda c}  \left( \vec{p} (\eta_i)  +  
\frac{\vec{A}(\eta) -
\vec{A} (\eta_i)}{2}\right) \cdot (\vec{A}(\eta) - \vec{A} (\eta_i))  \, , \\
\label{eom_en}
\varepsilon(\eta) & = \varepsilon(\eta_i) + \frac{\hat{\vec{k}}}{\lambda }  \left( \vec{p} (\eta_i)  +  
\frac{\vec{A}(\eta) -
\vec{A} (\eta_i)}{2}\right) \cdot (\vec{A}(\eta) - \vec{A} (\eta_i)) \, .
\eal

\bibliography{yakaboylu_bibliography}

\end{document}